\documentclass[12pt,english,aps,tightenlinesletterpaper,groupaddress,secnumarabic,nofootinbib]{revtex4}
\usepackage{mathptmx}

\usepackage[T1]{fontenc}
\usepackage[latin9]{inputenc}
\usepackage{babel}
\usepackage{amsmath}
\usepackage{amssymb}
\usepackage{graphicx}
\usepackage{esint}
\usepackage[unicode=true,pdfusetitle,
 bookmarks=true,bookmarksnumbered=false,bookmarksopen=false,
 breaklinks=false,pdfborder={0 0 1},backref=section,colorlinks=false]
 {hyperref}
\hypersetup{
 citecolor=blue}

\makeatletter

\providecommand{\tabularnewline}{\\}

\@ifundefined{textcolor}{}
{%
 \definecolor{BLACK}{gray}{0}
 \definecolor{WHITE}{gray}{1}
 \definecolor{RED}{rgb}{1,0,0}
 \definecolor{GREEN}{rgb}{0,1,0}
 \definecolor{BLUE}{rgb}{0,0,1}
 \definecolor{CYAN}{cmyk}{1,0,0,0}
 \definecolor{MAGENTA}{cmyk}{0,1,0,0}
 \definecolor{YELLOW}{cmyk}{0,0,1,0}
}

\usepackage{babel}

\usepackage{amsfonts}
\usepackage{bbm}
\usepackage{euscript}
\usepackage{dcolumn}
\usepackage{bm}
\usepackage{epsfig}\epsfclipon
\usepackage{slashed}

\newcommand{\cm}{c_-}
\newcommand{\cp}{c_+}
\newcommand{\Fa}{F_\mathrm{a}}
\newcommand{\Hin}{H}
\newcommand{\Mpl}{M_\mathrm{Pl}}
\newcommand{\niso}{n_\mathrm{I}}
\newcommand{\Pfit}{\Delta^2_\mathrm{fit}}
\newcommand{\Pm}{\tilde{\Phi}_-}
\newcommand{\Pp}{\tilde{\Phi}_+}
\newcommand{\Ppin}{\tilde{\Phi}_+(\eta_i)}
\newcommand{\Rinit}{\mathcal R}
\newcommand{\sci}[2]{#1$\times$10$^{#2}$}
\newcommand{\thp}{\theta_+(t_\mathrm{i})}

\usepackage{color}

\makeatother

\begin{document}

\title{A Bump in the Blue Axion Isocurvature Spectrum}

\author{Daniel J. H. Chung}

\email{danielchung@wisc.edu}

\affiliation{Department of Physics, University of Wisconsin-Madison, Madison,
WI 53706, USA}

\author{Amol Upadhye}

\email{aupadhye@wisc.edu}

\affiliation{Department of Physics, University of Wisconsin-Madison, Madison,
WI 53706, USA}
\begin{abstract}
Blue axion isocurvature perturbations are both theoretically well-motivated
and interesting from a detectability perspective. These power spectra
generically have a break from the blue region to a flat region. Previous
investigations of the power spectra were analytic, which left a gap
in the predicted spectrum in the break region due to the non-applicability
of the used analytic techniques. We therefore compute the isocurvature
spectrum numerically for an explicit supersymmetric axion model. We
find a bump that enhances the isocurvature signal for this class of
scenarios. A fitting function of three parameters is constructed that
fits the spectrum well for the particular axion model we study. This
fitting function should be useful for blue isocurvature signal hunting
in data and making experimental sensitivity forecasts. 
\end{abstract}
\maketitle

\section{Introduction}

In many proposals beyond the standard model (SM) of particle physics,
massive fields that live long enough to be dark matter candidates
commonly exist and could have been dynamical during inflation. The
well-known $\eta$-problem in inflation \cite{Copeland:1994vg} is
a statement of genericity of massive scalar fields with temporary
masses of order $H$ (expansion rate) during inflation for models
involving gravity (such as supergravity) \cite{Dine:1983ys,Bertolami:1987xb,Dine:1995uk}.
Some of these fields generically do not carry large energy density
during inflation (i.e.~they are not in the inflaton sector and are
often called spectators), and they will have de Sitter temperature
induced inhomogeneities which have a blue power spectrum due to the
field masses of order $H$ (see for example \cite{Linde:1996gt}).
If these fields are sufficiently secluded from both the inflaton sector
and the SM sector (\emph{i.e.} they are only very weakly interacting),
the blue spectrum will survive long enough for them to be observable
today \cite{Weinberg:2004kf} in the form of isocurvature perturbations
\cite{Takeuchi:2013hza,Dent:2012ne,Chluba:2013dna,Sekiguchi:2013lma}.%
\footnote{Non-Gaussianities discussed for example in \cite{Chen:2009zp,Craig:2014rta,Arkani-Hamed:2015bza,Dimastrogiovanni:2015pla}
do not necessarily require such weak interactions. %
} However, if the masses of order $H$ do not undergo a transition
to a different value at some point \emph{during} inflation, the energy
density dilution during inflation can make these noninflaton fields'
isocurvature perturbations nearly impossible to observe directly even
if they had a large amplitude blue isocurvature spectrum \cite{Chung:2015tha}.
Furthermore, a good theoretical motivation for such a dark matter
candidate is often desirable.

Axions \cite{Peccei:1977hh,Weinberg:1977ma,Wilczek:1977pj,Kim:1979if,Shifman:1979if,Zhitnitsky:1980tq,Dine:1981rt,Kim:2008hd},
which are well motivated from the perspective of solving the strong
CP problem, therefore are good candidates for generating blue isocurvature
perturbations \cite{Kasuya:2009up}. The current phenomenological
bounds require the axions to be very weakly interacting \cite{Raffelt:2006cw,Sikivie:2006ni,Kim:2008hd,Graham:2015ouw},
and therefore there is a phenomenological motivation for their seclusion
from the SM sector beyond the considerations of isocurvature perturbations.
Because they are pseudo-Nambu-Goldstone bosons, their coupling to
the inflaton can also naturally be limited, and therefore, one can
easily motivate their seclusion from the inflaton sector. Next comes
a most interesting ingredient. Even though axion masses are protected
by an anomalous global $U(1)_{PQ}$ (i.e. Peccei-Quinn symmetry denoted
as PQ symmetry), they would generically acquire masses of the order
of PQ order parameter field mass during inflation if the PQ order
parameter field is out of equilibrium and moving to its potential
minimum \cite{Chung:2015pga}. Through the $\eta$-problem mechanism
discussed above, PQ order parameter field can naturally have a mass
of order $H$, and therefore the axion temporarily has a mass of order
$H$ until the PQ order parameter reaches its minimum. Consequently,
the axion mass generically shuts off at some point during inflation,
allowing their energy density to survive inflationary dilution. During
the time when the axion mass has not shut off, axion quantum fluctuations
generate blue spectral inhomogeneities. Hence, axions possess all
the necessary ingredients for naturally generating an observable blue
isocurvature spectrum.

Blue axion isocurvature perturbations are therefore both theoretically
and observationally well-motivated. Previous investigations of the
power spectra were analytic \cite{Kasuya:2009up,Chung:2015pga}, leaving
a gap in the predicted primordial power spectrum near the spectral
scale where the axion mass turns off. In this work, we numerically
investigate this analytic gap region and find that there is a bump
in the power spectrum. This bump enhances the isocurvature amplitude
by an order unity factor, and such enhancements can facilitate the
observational detection or exclusion of this class of models. We construct
an economical fitting function Eq.~(\ref{eq:lessmodeldep}) consisting
of only 3 parameters. The fitting function will be useful in hunting
for such blue spectral signals in current and future observational
data. We also verify that the change in the isocurvature amplitude
after the end of inflation is of the expected negligible magnitude
of $(H/F_{a})^{2}\ll1$.

The order of presentation will be as follows, in the next section
we review the axion model of \cite{Kasuya:2009up} that we wish to
study in detail. In Sec.~\ref{sec:Setup-of-the}, we explain how
the numerical problem will be set up to deal with the issue of Planck
scale oscillation modes becoming light. In Sec.~\ref{sec:numerical_calculations},
we present the numerical computational results including the fitting
function. We conclude with a summary of the work.

\section{Axion model}

For concreteness of our numerical investigation, we consider the model
of \cite{Kasuya:2009up}. The qualitative features of this model are
expected to be generic, although quantitatively, the details may differ.

The authors of \cite{Kasuya:2009up} consider a supersymmetric axion
model with the following renormalizable superpotential 
\begin{equation}
W=h(\Phi_{+}\Phi_{-}-F_{a}^{2})\Phi_{0}
\end{equation}
where the subscripts on $\Phi$ indicate $U(1)_{PQ}$ global Peccei-Quinn
(PQ) charges. The F-term potential is 
\begin{equation}
V_{F}=h^{2}|\Phi_{+}\Phi_{-}-F_{a}^{2}|^{2}+h^{2}(|\Phi_{+}|^{2}+|\Phi_{-}|^{2})|\Phi_{0}|^{2}.
\end{equation}
A flat directions of $V_{F}$ exists along 
\begin{equation}
\Phi_{+}\Phi_{-}=F_{a}^{2}\,\,\,\,\,\,\,\,\,\,\,\,\Phi_{0}=0.\label{eq:flatdirection}
\end{equation}
The existence of this flat direction is important because this is
the reason why the effective PQ parameters will be rolling with a
mass of order $H$ during inflation (instead of being much heavier
and having already settled down), taking advantage of the inflationary
$\eta$-problem. Their low-scale SUSY-breaking terms are assumed to
be 
\begin{equation}
V_{{\rm soft}}=m_{+}^{2}|\Phi_{+}|^{2}+m_{-}^{2}|\Phi_{-}|^{2}+m_{0}^{2}|\Phi_{0}|^{2}
\end{equation}
where $m_{i}=O(\mbox{TeV})$. For most of the inflationary dynamics,
these parameters are irrelevant. The Kaehler potential induced scalar
potential is 
\begin{equation}
V_{K}=c_{+}H^{2}|\Phi_{+}|^{2}+c_{-}H^{2}|\Phi_{-}|^{2}+c_{0}H^{2}|\Phi_{0}|^{2}
\end{equation}
where $c_{+,-,0}$ are positive $O(1)$ constants. The parameter $c_{+}$
dominantly controls the blue spectral index. This setup implicitly
assumes that the inflaton sector can be arranged to have $H\ll F_{a}$
such that the flat directions are only lifted by the quadratic terms
at the renormalizable level.

Looking along the flat direction of Eq.~(\ref{eq:flatdirection}),
we set $\Phi_{0}=0$. The resulting relevant effective potential during
inflation is 
\begin{equation}
V\approx h^{2}|\Phi_{+}\Phi_{-}-F_{a}^{2}|^{2}+c_{+}H^{2}|\Phi_{+}|^{2}+c_{-}H^{2}|\Phi_{-}|^{2}.
\end{equation}
During inflation, the minimum of $\Phi_{\pm}$ lies at 
\begin{equation}
|\Phi_{\pm}^{\mbox{min}}|\approx\left(\frac{c_{\mp}}{c_{\pm}}\right)^{1/4}F_{a}.
\end{equation}
The key initial condition is that $\Phi_{\pm}$ starts out away from
the minimum with a magnitude much larger than $O(F_{a})$ and rolls
towards the minimum during inflation. This implies the $U(1)_{PQ}$
symmetry is broken during inflation. Hence, there will be a linear
combination of the phases of $\Phi_{\pm}$ which will be the Nambu-Goldstone
boson associated with the broken $U(1)_{PQ}$. In particular, with
the parameterization

\begin{equation}
\Phi_{\pm}\equiv\frac{\varphi_{\pm}}{\sqrt{2}}\exp\left(i\frac{a_{\pm}}{\sqrt{2}\varphi_{\pm}}\right)\label{eq:angularparam}
\end{equation}
where $\varphi_{\pm}$ and $a_{\pm}$ are real, the axion is 
\begin{equation}
a=\frac{\varphi_{+}}{\sqrt{\varphi_{+}^{2}+\varphi_{-}^{2}}}a_{+}-\frac{\varphi_{-}}{\sqrt{\varphi_{+}^{2}+\varphi_{-}^{2}}}a_{-}
\end{equation}
while the heavier combination 
\begin{equation}
b=\frac{\varphi_{-}}{\sqrt{\varphi_{+}^{2}+\varphi_{-}^{2}}}a_{+}+\frac{\varphi_{+}}{\sqrt{\varphi_{+}^{2}+\varphi_{-}^{2}}}a_{-}\label{eq:biszero-1}
\end{equation}
is governed by the potential 
\begin{equation}
V_{b}=-h^{2}F_{a}^{2}\varphi_{+}\varphi_{-}\cos\left(\frac{\sqrt{\varphi_{+}^{2}+\varphi_{-}^{2}}}{\varphi_{+}\varphi_{-}}b\right).\label{eq:bheavy}
\end{equation}
Since the $b$ field is heavy (\emph{i.e.} $(\varphi_{+}^{2}+\varphi_{-}^{2})F_{a}^{2}/(\varphi_{+}\varphi_{-})\gg H^{2}$),
it is not dynamically important. Hence, one can gain some intuition
for how the axion composition time evolves by setting $b=0$. When
$\varphi_{+}$ is large, the axion is dominantly $a_{+}$ and later
when $\varphi_{+}$ becomes comparable to $\varphi_{-}$, the axion
is a mixture of $a_{-}$ and $a_{+}$.

\section{\label{sec:Setup-of-the}Setup of the numerical problem}

Here we present a semi-numerical approach to the mode problem. In
setting up the numerical problem, the equation of motion in terms
of $a$ field is far more complicated than in the $\{\Phi_{\pm},\Phi_{0}\}$
basis. We will therefore set up the spectral numerical computation
in terms of $\{\Phi_{\pm},\Phi_{0}\}$. The background equations are
\begin{equation}
\ddot{\Phi}_{+}+3H\dot{\Phi}_{+}+(c_{+}H^{2}+m_{+}^{2})\Phi_{+}+h^{2}(\Phi_{+}\Phi_{-}-F_{a}^{2})\Phi_{-}^{*}+h^{2}\Phi_{+}|\Phi_{0}|^{2}=0
\end{equation}
\begin{equation}
\ddot{\Phi}_{-}+3H\dot{\Phi}_{-}+(c_{-}H^{2}+m_{-}^{2})\Phi_{-}+h^{2}(\Phi_{+}\Phi_{-}-F_{a}^{2})\Phi_{+}^{*}+h^{2}\Phi_{-}|\Phi_{0}|^{2}=0
\end{equation}
\begin{equation}
\ddot{\Phi}_{0}+3H\dot{\Phi}_{0}+(c_{0}H^{2}+m_{0}^{2})\Phi_{0}+h^{2}(|\Phi_{+}|^{2}+|\Phi_{-}|^{2})\Phi_{0}=0
\end{equation}
and the fluctuation equations in Fourier space are 
\begin{eqnarray}
\delta\ddot{\Phi}_{+}+3H\delta\dot{\Phi}_{+}+(c_{+}H^{2}+m_{+}^{2}+\frac{k^{2}}{a^{2}})\delta\Phi_{+}+h^{2}(\Phi_{+}\Phi_{-}-F_{a}^{2})\delta\Phi_{-}^{*}+h^{2}\delta\Phi_{+}|\Phi_{0}|^{2}\nonumber \\
+h^{2}(\Phi_{+}\delta\Phi_{-})\Phi_{-}^{*}+h^{2}(\delta\Phi_{+}\Phi_{-})\Phi_{-}^{*}+h^{2}\Phi_{+}\delta\Phi_{0}\Phi_{0}^{*}+h^{2}\Phi_{+}\Phi_{0}\delta\Phi_{0}^{*} & = & 0
\end{eqnarray}
\begin{eqnarray}
\delta\ddot{\Phi}_{-}+3H\delta\dot{\Phi}_{-}+(c_{-}H^{2}+m_{-}^{2}+\frac{k^{2}}{a^{2}})\delta\Phi_{-}+h^{2}(\Phi_{+}\Phi_{-}-F_{a}^{2})\delta\Phi_{+}^{*}+h^{2}\delta\Phi_{-}|\Phi_{0}|^{2}\nonumber \\
+h^{2}(\delta\Phi_{+}\Phi_{-})\Phi_{+}^{*}+h^{2}(\Phi_{+}\delta\Phi_{-})\Phi_{+}^{*}+h^{2}\Phi_{-}\delta\Phi_{0}\Phi_{0}^{*}+h^{2}\Phi_{-}\Phi_{0}\delta\Phi_{0}^{*} & = & 0
\end{eqnarray}
\begin{eqnarray}
\delta\ddot{\Phi}_{0}+3H\delta\dot{\Phi}_{0}+(c_{0}H^{2}+m_{0}^{2}+\frac{k^{2}}{a^{2}})\delta\Phi_{0}+h^{2}(|\Phi_{+}|^{2}+|\Phi_{-}|^{2})\delta\Phi_{0}\nonumber \\
+h^{2}(\Phi_{+}\delta\Phi_{+}^{*}+\delta\Phi_{+}\Phi_{+}^{*}+\delta\Phi_{-}\Phi_{-}^{*}+\Phi_{-}\delta\Phi_{-}^{*})\Phi_{0} & = & 0.
\end{eqnarray}
Since we are interested in flat direction solutions to the background
equations and since the $\Phi_{0}$ mass is extremely large for the
large displacements of $\Phi_{+}$ that we are interested in, we restrict
ourselves to the 
\begin{equation}
\Phi_{0}=0
\end{equation}
solution to the equations of motion. This simplifies the perturbation
equations significantly. Furthermore, we can rephase the the background
fields such that only real background functions need to be evolved
because of the CP symmetry of the background equations. This means
that if we set the initial condition such that 
\begin{equation}
\tilde{\Phi}_{\pm}=\Phi_{\pm}\exp\left(\mp i\thp\right)\label{eq:phaserotation}
\end{equation}
where $\tilde{\Phi}_{\pm}$ is real, $\tilde{\Phi}_{\pm}$ will remain
real. Note that the opposite rephasing of $\Phi_{\pm}$ initial conditions
is consistent with the heavy mode $b=0$ in the background equation
(see Eq.~(\ref{eq:bheavy})). The initial time $t_{i}$ must be chosen
such that the longest observable wave vector $k_{{\rm min}}$ must
be be subhorizon at time $t_{i}$.

For quantization of the fluctuations, it is also convenient to decompose
the perturbations into real scalar fields $\{R_{\pm},I_{\pm}\}$:
\begin{equation}
\delta\Phi_{\pm}=R_{\pm}+iI_{\pm}
\end{equation}
\begin{equation}
\delta\Phi_{0}=Z_{r}+iZ_{i}.
\end{equation}
The quantum mode equations%
\footnote{For example, we can write 
\begin{equation}
\Re\left(\Phi_{+}(t,\vec{x})\right)=\tilde{\Phi}_{+}(t)\cos(\theta_{+}(t_{i}))+\int\frac{d^{3}k}{(2\pi)^{3/2}}\left[a_{\vec{k}}^{(R+)}R_{+}e^{i\vec{k}\cdot\vec{x}}+a_{\vec{k}}^{(R+)\dagger}R_{+}^{*}e^{-i\vec{k}\cdot\vec{x}}\right]
\end{equation}
\begin{equation}
\Im\left(\Phi_{+}(t,\vec{x})\right)=\tilde{\Phi}_{+}(t)\sin(\theta_{+}(t_{i}))+\int\frac{d^{3}k}{(2\pi)^{3/2}}\left[a_{\vec{k}}^{(I+)}I_{+}e^{i\vec{k}\cdot\vec{x}}+a_{\vec{k}}^{(I+)\dagger}I_{+}^{*}e^{-i\vec{k}\cdot\vec{x}}\right]
\end{equation}
in a creation-annihilation operator expansion where $\{R_{+},I_{+}\}$
are mode functions satisfying mode equations.%
} therefore are 
\begin{eqnarray}
\ddot{R}_{+}+3H\dot{R}_{+}+(c_{+}H^{2}+m_{+}^{2}+\frac{k^{2}}{a^{2}})R_{+}+h^{2}(\tilde{\Phi}_{+}\tilde{\Phi}_{-}-F_{a}^{2})R_{-}\nonumber \\
+h^{2}[\cos\left(2\theta_{+}(t_{i})\right)R_{-}-\sin\left(2\theta_{+}(t_{i})\right)I_{-}]\tilde{\Phi}_{+}\tilde{\Phi}_{-}+h^{2}\tilde{\Phi}_{-}^{2}R_{+} & = & 0\label{eq:rplusmode}
\end{eqnarray}
\begin{eqnarray}
\ddot{I}_{+}+3H\dot{I}_{+}+(c_{+}H^{2}+m_{+}^{2}+\frac{k^{2}}{a^{2}})I_{+}-h^{2}(\tilde{\Phi}_{+}\tilde{\Phi}_{-}-F_{a}^{2})I_{-}\nonumber \\
+h^{2}\left[\cos\left(2\theta_{+}(t_{i})\right)I_{-}+\sin\left(2\theta_{+}(t_{i})\right)R_{-}\right]\tilde{\Phi}_{+}\tilde{\Phi}_{-}+h^{2}\tilde{\Phi}_{-}^{2}I_{+} & = & 0
\end{eqnarray}
\begin{eqnarray}
\ddot{R}_{-}+3H\dot{R}_{-}+(c_{-}H^{2}+m_{-}^{2}+\frac{k^{2}}{a^{2}})R_{-}+h^{2}(\tilde{\Phi}_{+}\tilde{\Phi}_{-}-F_{a}^{2})R_{+}\nonumber \\
+h^{2}\left(\cos\left(2\theta_{+}(t_{i})\right)R_{+}+\sin\left(2\theta_{+}(t_{i})\right)I_{+}\right)\tilde{\Phi}_{+}\tilde{\Phi}_{-}+h^{2}\tilde{\Phi}_{+}^{2}R_{-} & = & 0
\end{eqnarray}
\begin{eqnarray}
\ddot{I}_{-}+3H\dot{I}_{-}+(c_{-}H^{2}+m_{-}^{2}+\frac{k^{2}}{a^{2}})I_{-}-h^{2}(\tilde{\Phi}_{+}\tilde{\Phi}_{-}-F_{a}^{2})I_{+}\nonumber \\
+h^{2}\left(\cos\left(2\theta_{+}(t_{i})\right)I_{+}-\sin\left(2\theta_{+}(t_{i})\right)R_{+}\right)\tilde{\Phi}_{+}\tilde{\Phi}_{-}+h^{2}\tilde{\Phi}_{+}^{2}I_{-} & = & 0
\end{eqnarray}
\begin{equation}
\ddot{Z}_{r,i}+3H\dot{Z}_{r,i}+(c_{0}H^{2}+m_{0}^{2}+\frac{k^{2}}{a^{2}}+h^{2}[|\tilde{\Phi}_{+}|^{2}+|\tilde{\Phi}_{-}|^{2}])Z_{r,i}=0.\label{eq:Zmode}
\end{equation}
Note that $Z_{r,i}$ modes are completely decoupled from the other
modes.

For $\theta_{+}(t_{i})\ll1$, the axion correlator that we are interested
in computing is 
\begin{eqnarray}
\left\langle \frac{\delta a}{a}\frac{\delta a}{a}\right\rangle \mathcal{D} & = & \langle I_{-}I_{-}\rangle\tilde{\Phi}_{-}^{2}+\langle I_{+}I_{+}\rangle\tilde{\Phi}_{+}^{2}-\left[\langle I_{+}I_{-}\rangle+\langle I_{-}I_{+}\rangle\right]\tilde{\Phi}_{-}\tilde{\Phi}_{+}-\left[\langle I_{-}R_{-}\rangle+\langle R_{-}I_{-}\rangle\right]\tilde{\Phi}_{-}^{2}\theta_{+}(t_{i})\nonumber \\
 &  & -\left[\langle I_{-}R_{+}\rangle+\langle R_{+}I_{-}\rangle\right]\tilde{\Phi}_{-}\tilde{\Phi}_{+}\theta_{+}(t_{i})+\left[\langle I_{+}R_{-}\rangle+\langle R_{-}I_{+}\rangle\right]\tilde{\Phi}_{-}\tilde{\Phi}_{+}\theta_{+}(t_{i})\nonumber \\
 &  & +\left[\langle I_{+}R_{+}\rangle+\langle R_{+}I_{+}\rangle\right]\tilde{\Phi}_{+}^{2}\theta_{+}(t_{i})\label{eq:finalcorrelator}
\end{eqnarray}
where 
\begin{eqnarray}
\mathcal{D} & \equiv & \left(\tilde{\Phi}_{+}^{2}+\tilde{\Phi}_{-}^{2}\right)^{2}\theta_{+}^{2}(t_{i}),
\end{eqnarray}
and here we follow the typical abuse of notation in which the quantum
fields and their mode functions are denoted with the same symbols.
This correlator is related to the primordial isocurvature spectrum through
the equation 
\begin{equation}
\Delta_{S}^{2}(t,\vec{p})=4\omega_{a}^{2}\frac{p^{3}}{2\pi^{2}}\int\frac{d^{3}q}{(2\pi)^{3}}\left\langle \frac{\delta a(t,\vec{p})}{a}\frac{\delta a(t,\vec{q})}{a}\right\rangle \label{eq:isocurvatureexpl}
\end{equation}
where we use the ratio of a common formula for a QCD axion energy
density to cold dark matter energy density (e.g. equation 14 of \cite{Kawasaki:2013ae})
\begin{eqnarray}
\omega_{a} & \equiv & \frac{\Omega_{a}}{\Omega_{{\rm cdm}}}\label{eq:darkmatterfraction}\\
 & = & W_{a}\theta_{+}^{2}(t_{i})\left(\frac{\sqrt{2}\left(\tilde{\Phi}_{+}^{2}(t_{f})+\tilde{\Phi}_{-}^{2}(t_{f})\right)^{1/2}}{10^{12}{\rm GeV}}\right)^{n_{PT}}
\end{eqnarray}
where $W_{a}\approx1.5$ and $n_{PT}\approx1.19$ and $t_{f}$ is
the time just before the QCD phase transition.%
\footnote{The fields $\tilde{\Phi}_{\pm}$ have settled down long before this.%
} This formula differs from equation 203 of \cite{Chung:2015pga} by
a factor of $2^{n_{PT}/2}$ because of a mistake in the way the axion
decay constant was defined there. Here we are defining the effective
decay constant as 
\begin{equation}
f_{a}=\sqrt{2}\left(\tilde{\Phi}_{+}^{2}(t_{f})+\tilde{\Phi}_{-}^{2}(t_{f})\right)^{1/2}.
\end{equation}
Each of the correlators in Eq.~(\ref{eq:finalcorrelator}) is computed
using the procedure as specified in \cite{Kawasaki:2013ae}. For example,
the $I_{+}$ correlator is computed as 
\begin{equation}
\langle I_{+}I_{+}\rangle\rightarrow\langle I_{+}(t,\vec{p})I_{+}(t,\vec{q})\rangle=(2\pi)^{3}\sum_{\alpha=1}^{6}I_{+}^{(\alpha)}(t,\vec{p})I_{+}^{(\alpha)*}(t,\vec{q})\delta^{(3)}(\vec{q}+\vec{p})\label{eq:correlator}
\end{equation}
where the $\alpha$-labeled boundary conditions will be discussed
shortly (around Eq.~(\ref{eq:boundarycond})). There we will explain
how $\alpha\in\{5,6\}$ will not contribute since those modes do not
mix. The goal of the rest of this section is to set up the numerical
problem to compute $\Delta_{S}^{2}(t,\vec{p})/\omega_{a}^{2}$ to
about 20\% accuracy after the end of inflation for a wide range of
parameters which are $\{c_{+},\,\, c_{-},\,\, F_{a},\,\, H,\,\,\tilde{\Phi}_{+}(t_{i}),\,\,\theta(t_{i}),\,\, h\}$
where $t_{i}$ is the initial time during inflation when the effective
field theory describing this axion model is valid.%
\footnote{Also, for the target level of accuracy, we can ignore slow-roll evolution
of the expansion rate during inflation and set $H$ to be a constant.%
} Although the accuracy goal may naively seem poor, it is actually
only modestly larger than the Planck bound since a 20\% accuracy and
an isocurvature fraction of about 10\% implies a few percent accuracy
in the total power spectrum. Note $m_{\pm}\ll H$ is assumed such
that they are not relevant for this computation. Also, note that we
do not need to compute $Z_{r,i}$ since they do not mix with $\{R_{\pm},I_{\pm}\}$
and do not enter in Eq.~(\ref{eq:finalcorrelator}). We will henceforth
drop any discussion of $Z_{r,i}$.

\subsection{Ideal initial conditions}

As noted in \cite{Chen:2015dga}, to stay consistent with the tree-level
truncation of the in-in formalism, we should set one of $\psi\equiv(R_{+},I_{+},R_{-},I_{-})$
modes initially non-zero and every other mode initially zero. The
non-zero boundary condition should be adiabatic (which is approximately
equivalent to Bunch-Davies vacuum). For example, the boundary conditions
at a time when $k\gg a(t_{i\,{\rm num}})H$ can be taken to be (for
the numerical run $\alpha$) 
\begin{equation}
\psi_{l}^{(\alpha)}|_{t=t_{i\,{\rm num}}(k)}=(U^{\dagger})_{l\alpha}\frac{N_{\alpha}}{\sqrt{2}\lambda_{\alpha}^{1/4}a^{3/2}(t_{i\,{\rm num}}(k))}\label{eq:boundarycond}
\end{equation}
\begin{equation}
\frac{d}{dt}\psi_{l}^{(\alpha)}|_{t=t_{i\,{\rm num}}(k)}=-i(U^{\dagger})_{l\alpha}\frac{N_{\alpha}\lambda_{\alpha}^{1/4}}{\sqrt{2}a^{3/2}(t_{i\,{\rm num}}(k))}\label{eq:boundarycond2}
\end{equation}
where 
\begin{equation}
N_{\alpha}=\frac{1}{\sqrt{2}}\label{eq:normalization}
\end{equation}
is a normalization factor that can come from non-canonical normalization
of the kinetic term, $U$ is a mixing matrix that diagonalizes the
dispersion squared matrix, and we are assuming that the mixing matrix
time derivative is negligible at the time of the initial conditions.
Note the minus sign on the ``$i$'' corresponds to defining the
positive frequency modes. However, this procedure is numerically expensive
and impractical since some of the eigenvalues start close to Planckian
mass values and the oscillations need to be tracked until the time
when the mass scales reach $10^{11}$ GeV.

\subsection{Semi-numerical WKB approach and boundary conditions for full numerics}

For long wavelength modes, $k/(aH)\gg1$ corresponds to times when
$\tilde{\Phi}_{+}(t)\sim O(M_{p})$. This means that some of the long
wavelength modes in $\psi=(R_{+},I_{+},R_{-},I_{-})$ have Planckian
masses during this time and one might naively set boundary conditions
for modes when the oscillation frequency is of order the Planck scale.
Because Planck scale frequency oscillations are physically irrelevant
for our observables, such modes can be integrated out. However, the
masses eventually change as a function of time such that these modes
become relevant. For the numerical approach, we need a prescription
to set the boundary conditions for the shorter wavelength modes whose
oscillation frequency is always small consistently with the longer
wavelength modes which start with Planckian oscillations and become
non-Planckian. We construct below a consistent prescription that can
be described as follows. For long wavelength heavy modes, we use analytic
WKB solutions which are accurate. For lighter long wavelength modes,
we take a semi-numerical modified WKB approach to solve the modes
accurately in possible turning point regions. When the WKB and the
modified WKB solutions begin to depart from being excellent approximations,
all the masses are of order $F_{a}\ll M_{p}$ or smaller, and we can
during this period safely compute all the modes numerically without
the expense of computing irrelevant fast oscillations.

First, let us set up the math problem explicitly. As noted before,
because $\delta\Phi_{0}$ oscillations do not contribute to the tree-level
correlator we wish to compute, we can reduce the numerical problem
to 4 real quantum fields containing 4 independent complex modes.%
\footnote{The modes $Z_{R,i}$ do not mix with the rest of the modes in Eqs.~(\ref{eq:rplusmode})
through (\ref{eq:Zmode}). %
} For the numerical study which aims for an accuracy of about 20\%,
we set $H=$constant during inflation%
\footnote{Secular effects due to time evolving $H$ give a correction of order
$10\epsilon$ during the $O(10)$ e-folds of inflationary phase that
is observable.%
} and define the $\psi$ vector to be 
\begin{equation}
\psi\equiv(R_{+},I_{+},R_{-},I_{-})
\end{equation}
to write the complex mode equations as 
\begin{equation}
\psi''+3\psi'+M^{2}\psi=0\label{eq:originalmodeeq}
\end{equation}
where 
\begin{equation}
M^{2}\equiv\left(\begin{array}{cc}
\frac{\mu_{+}^{2}}{H^{2}}I & \frac{\mu_{-}^{2}}{H^{2}}Q^{(+)}\\
\frac{\mu_{-}^{2}}{H^{2}}Q^{(-)} & \frac{\mu_{-}^{2}}{H^{2}}I
\end{array}\right)
\end{equation}
\begin{equation}
Q^{(\pm)}=\frac{h^{2}\tilde{\Phi}_{+}\tilde{\Phi}_{-}}{\mu_{-}^{2}}\left[I+\left(\begin{array}{cc}
\cos2\theta-1 & \mp\sin2\theta\\
\pm\sin2\theta & \cos2\theta-1
\end{array}\right)\right]+\frac{h^{2}}{\mu_{-}^{2}}(\tilde{\Phi}_{+}\tilde{\Phi}_{-}-F_{a}^{2})\sigma_{\alpha\beta}^{(3)}\label{eq:mixingmatrixeqlam2}
\end{equation}
\begin{equation}
\mu_{\pm}^{2}=h^{2}\tilde{\Phi}_{\mp}^{2}+c_{\pm}H^{2}+m_{\pm}^{2}+\frac{k^{2}}{a^{2}}
\end{equation}
\begin{equation}
\sigma^{(3)}=\left(\begin{array}{cc}
1 & 0\\
0 & -1
\end{array}\right)
\end{equation}
where $m_{\pm}^{2}$ are small quantities irrelevant for leading approximation
cosmology. We want to solve this problem following the mode from the
subhorizon period until past the end of inflation. During this entire
period of interest, there are subhorizon WKB approximate oscillations,
horizon-crossing possibly involving turning points for light modes,
and a nonadiabatic period when WKB approximation breaks down and one
must solve the mode equations fully numerically. The utility of the
modified WKB approach below will be that it will set up a numerical
problem that smoothly connects the WKB and the turning point regions
into a differential equation for a single complex function (not a
vector of complex functions) without necessitating the definition
of an arbitrary turning point for light modes.

To diagonalize the mass squared matrix, we rewrite it as follows:
\begin{equation}
M^{2}=M_{0}^{2}+\lambda_{21}\Delta_{21}+\lambda_{22}\Delta_{22}+\lambda_{33}\Delta_{3}
\end{equation}
where 
\begin{equation}
\Delta_{21}=\left(\begin{array}{cc}
0 & \,\,\,\,\,\,\,\, I\\
I & \,\,\,\,\,\,\,\,0
\end{array}\right)
\end{equation}
\begin{equation}
\lambda_{21}\equiv\frac{h^{2}\left(\cos2\theta-1\right)\Phi_{+}\Phi_{-}}{H^{2}}
\end{equation}
\begin{equation}
\Delta_{22}=\left(\begin{array}{cc}
0 & -i\sigma_{\alpha\beta}^{(2)}\\
i\sigma_{\alpha\beta}^{(2)} & 0
\end{array}\right)
\end{equation}
\begin{equation}
\lambda_{22}\equiv\frac{h^{2}\Phi_{+}\Phi_{-}\sin2\theta}{H^{2}}
\end{equation}
\begin{equation}
\Delta_{3}=\left(\begin{array}{cc}
0 & \sigma_{\alpha\beta}^{(3)}\\
\sigma_{\alpha\beta}^{(3)} & 0
\end{array}\right)
\end{equation}
\begin{equation}
\lambda_{33}\equiv\frac{h^{2}}{H^{2}}(\Phi_{+}\Phi_{-}-F_{a}^{2})
\end{equation}
\begin{equation}
\sigma^{(2)}=\left(\begin{array}{cc}
0 & \,\,\,\,\,\,\,\,-i\\
i & \,\,\,\,\,\,\,\,0
\end{array}\right).
\end{equation}
The first eigenvalue is 
\begin{equation}
E_{1}^{2}=\frac{W_{D}^{2}-\sqrt{4F_{2}^{4}+W_{-}^{4}-4\lambda_{33}^{2}H^{4}\left(\frac{2F_{2}^{2}}{\lambda_{33}H^{2}}-1\right)}}{2H^{2}}\label{eq:starteigs}
\end{equation}
and the corresponding eigenvector is 
\begin{equation}
|E_{1}^{2}\rangle=\mathcal{N}_{1}(a_{1},b_{1},c_{1},1)\,\,\,\,\,\,\,\,\,\,\,\,\,\mathcal{N}_{1}\equiv\left(a_{1}^{2}+b_{1}^{2}+c_{1}^{2}+1\right)^{-1/2}
\end{equation}
\begin{eqnarray}
a_{1} & = & \frac{\left(F_{2}^{2}-\tilde{F}_{a}^{2}h^{2}\right)}{2\lambda_{22}H^{2}}\left(\sqrt{4+\frac{W_{-}^{4}}{\left(F_{2}^{2}-\lambda_{33}H^{2}\right)^{2}}}+\frac{W_{-}^{2}}{F_{2}^{2}-\lambda_{33}H^{2}}\right)\\
 & \approx & \frac{\lambda_{22}H^{2}}{2h^{2}\tilde{F}_{a}^{2}}\left(\sqrt{1+\frac{W_{-}^{4}}{4\left(F_{2}^{2}-\lambda_{33}H^{2}\right)^{2}}}+\frac{W_{-}^{2}}{2\left(F_{2}^{2}-\lambda_{33}H^{2}\right)}\right)
\end{eqnarray}
\begin{equation}
b_{1}=-\sqrt{1+\frac{W_{-}^{4}}{4\left(F_{2}^{2}-\lambda_{33}H^{2}\right)^{2}}}-\frac{W_{-}^{2}}{2\left(F_{2}^{2}-\lambda_{33}H^{2}\right)}
\end{equation}
\begin{eqnarray}
c_{1} & = & \frac{F_{2}^{2}-\tilde{F}_{a}^{2}h^{2}}{\lambda_{22}H^{2}}\\
 & \approx & \frac{\lambda_{22}H^{2}}{2h^{2}\tilde{F}_{a}^{2}}
\end{eqnarray}
\begin{equation}
F_{2}^{4}\equiv h^{4}\tilde{F}_{a}^{4}+\lambda_{22}^{2}H^{4}
\end{equation}
\begin{equation}
W_{D}^{2}\equiv h^{2}[\Phi_{+}^{2}+\Phi_{-}^{2}]+(c_{+}+c_{-})H^{2}+2\frac{k^{2}}{a^{2}}
\end{equation}
\begin{equation}
W_{-}^{2}\equiv h^{2}(\Phi_{+}^{2}-\Phi_{-}^{2})+(c_{-}-c_{+})H^{2}
\end{equation}
\begin{equation}
\tilde{F}_{a}^{2}\equiv F_{a}^{2}+\frac{\lambda_{21}H^{2}}{h^{2}}\label{eq:replacement}
\end{equation}
where we will generally denote unit normalization factors as $\mathcal{N}_{i}$
for the $i$th eigenvector and we assumed 
\begin{equation}
F_{2}^{2}-\lambda_{33}H^{2}>0.
\end{equation}

The next eigenvalue and eigenvector can be written similarly: 
\begin{equation}
E_{2}^{2}=\frac{W_{D}^{2}+\sqrt{4F_{2}^{4}+W_{-}^{4}-4\lambda_{33}^{2}H^{4}(\frac{2F_{2}^{2}}{\lambda_{33}H^{2}}-1)}}{2H^{2}}
\end{equation}
\begin{equation}
|E_{2}^{2}\rangle=\mathcal{N}_{2}(a_{2},b_{2},c_{2},1)
\end{equation}
\begin{eqnarray}
a_{2} & = & \frac{\left(F_{2}^{2}-\tilde{F}_{a}^{2}h^{2}\right)}{2\lambda_{22}H^{2}}\left(-\sqrt{4+\frac{W_{-}^{4}}{\left(F_{2}^{2}-\lambda_{33}H^{2}\right)^{2}}}+\frac{W_{-}^{2}}{F_{2}^{2}-\lambda_{33}H^{2}}\right)\\
 & \approx & \frac{\lambda_{22}H^{2}}{4h^{2}\tilde{F}_{a}^{2}}\left(-\sqrt{4+\frac{W_{-}^{4}}{\left(F_{2}^{2}-\lambda_{33}H^{2}\right)^{2}}}+\frac{W_{-}^{2}}{F_{2}^{2}-\lambda_{33}H^{2}}\right)
\end{eqnarray}
\begin{equation}
b_{2}=\sqrt{1+\frac{W_{-}^{4}}{4\left(F_{2}^{2}-\lambda_{33}H^{2}\right)^{2}}}-\frac{W_{-}^{2}}{2\left(F_{2}^{2}-\lambda_{33}H^{2}\right)}
\end{equation}
\begin{equation}
c_{2}=c_{1}.
\end{equation}
The next eigenvector is interesting because one of the coefficients
have a large correction with respect to the eigenvector that is obtained
with $\lambda_{22}=0$: 
\begin{equation}
E_{3}^{2}=\frac{W_{D}^{2}-\sqrt{4F_{2}^{4}+W_{-}^{4}+4\lambda_{33}^{2}H^{4}\left(\frac{2F_{2}^{2}}{\lambda_{33}H^{2}}+1\right)}}{2H^{2}}\label{eq:e3sqeig}
\end{equation}
\begin{equation}
|E_{3}^{2}\rangle=\mathcal{N}_{3}(a_{3},b_{3},c_{3},\lambda_{22})
\end{equation}
\begin{equation}
a_{3}=\frac{\left(F_{2}^{2}+\tilde{F}_{a}^{2}h^{2}\right)}{2H^{2}}\left(\sqrt{4+\frac{W_{-}^{4}}{\left(F_{2}^{2}+\lambda_{33}H^{2}\right)^{2}}}+\frac{W_{-}^{2}}{F_{2}^{2}+\lambda_{33}H^{2}}\right)
\end{equation}
\begin{equation}
b_{3}=\lambda_{22}\left[\sqrt{1+\frac{W_{-}^{4}}{4\left(F_{2}^{2}+\lambda_{33}H^{2}\right)^{2}}}+\frac{W_{-}^{2}}{2\left(F_{2}^{2}+\lambda_{33}H^{2}\right)}\right]\label{eq:becorrection}
\end{equation}
\begin{equation}
c_{3}=-\frac{F_{2}^{2}+\tilde{F}_{a}^{2}h^{2}}{H^{2}}.\label{eq:c3correction}
\end{equation}
The large $\lambda_{22}$ effect can be attributed to the fact that
although $\lambda_{33}$ is the perturbation that breaks the degeneracy,
the eigenvectors with $\lambda_{22}$ turned off already diagonalizes
the perturbation matrix. On the other hand, when both $\lambda_{22}$
and $\lambda_{33}$ turn on, there is an off-diagonal matrix element
of the $\lambda_{33}$ perturbation in the original basis. This leads
to a large degenerate perturbation theory correction. For example,
$b_{3}$ is a $\lambda_{22}\lambda_{33}/\lambda_{33}$ effect. The
$\lambda_{33}/\lambda_{33}$ is typical of degenerate perturbation
theory effect, but $\lambda_{22}$ multiplying it shows that this
is actually a second order effect in the perturbation. The fourth
eigenvector system is characterized by 
\begin{equation}
E_{4}^{2}=\frac{W_{D}^{2}+\sqrt{4F_{2}^{4}+W_{-}^{4}+4\lambda_{33}^{2}H^{4}\left(\frac{2F_{2}^{2}}{\lambda_{33}H^{2}}+1\right)}}{2H^{2}}
\end{equation}
\begin{equation}
|E_{4}^{2}\rangle=\mathcal{N}_{4}(a_{4},b_{4},c_{4},\lambda_{22})
\end{equation}
\begin{equation}
a_{4}=\frac{\left(F_{2}^{2}+\tilde{F}_{a}^{2}h^{2}\right)}{2H^{2}}\left(-\sqrt{4+\frac{W_{-}^{4}}{\left(F_{2}^{2}+\lambda_{33}H^{2}\right)^{2}}}+\frac{W_{-}^{2}}{F_{2}^{2}+\lambda_{33}H^{2}}\right)
\end{equation}

\begin{equation}
b_{4}=\lambda_{22}\left[-\sqrt{1+\frac{W_{-}^{4}}{4\left(F_{2}^{2}+\lambda_{33}H^{2}\right)^{2}}}+\frac{W_{-}^{2}}{2\left(F_{2}^{2}+\lambda_{33}H^{2}\right)}\right]
\end{equation}
\begin{equation}
c_{4}=c_{3}\label{eq:endeigs}
\end{equation}

We can obtain some intuition about this eigensystem if we set $\lambda_{21}=\lambda_{22}=\lambda_{33}=0$
and keep the leading eigenvectors and their oscillation frequencies
in the limit of no $\theta$ induced mixing (i.e. $\theta F_{a}^{2}/H^{2}\rightarrow0$
and $H^{2}/F_{a}^{2}\rightarrow0$) and $\{F_{a}^{2}/\Phi_{+}^{2}\rightarrow0,\,\, h^{2}\Phi_{-}^{2}/H^{2}\rightarrow0\}$:
\begin{equation}
|E_{1}^{2}\rightarrow k^{2}/(aH)^{2}\rangle\rightarrow(0,1,0,0)\label{eq:firstapproximateeigenvector}
\end{equation}
\begin{equation}
|E_{2}^{2}\rightarrow h^{2}\Phi_{+}^{2}/H^{2}\rangle\rightarrow(0,0,0,1)
\end{equation}
\begin{equation}
|E_{3}^{2}\rightarrow k^{2}/(aH)^{2}\rangle\rightarrow(1,0,0,0)\label{eq:naieveigenvector1}
\end{equation}
\begin{equation}
|E_{4}^{2}\rightarrow h^{2}\Phi_{+}^{2}/H^{2}\rangle\rightarrow(0,0,1,0)\label{eq:naiveeigenvectors}
\end{equation}
where we have chosen the normalization factors $\mathcal{N}_{i}$
such that the largest component of the eigenvector is positive. The
ideal initial conditions discussed abstractly in Eqs.~(\ref{eq:boundarycond})
and (\ref{eq:boundarycond2}) are

\begin{equation}
\psi^{(j)}(\eta_{i})2\sqrt{E_{j}(\eta_{i})H(\eta_{i})}a^{3/2}(\eta_{i})=|E_{j}^{2}(\eta_{i})\rangle2\sqrt{E_{j}(\eta_{i})H(\eta_{i})}a^{3/2}(\eta_{i})\approx\mathcal{N}_{j}\left(\begin{array}{c}
a_{j}\\
b_{j}\\
c_{j}\\
d_{j}
\end{array}\right)\label{eq:valbc}
\end{equation}
\begin{eqnarray}
\partial_{\eta}\psi^{(j)}(\eta_{i})2\sqrt{E_{j}(\eta_{i})H(\eta_{i})}a^{3/2}(\eta_{i}) & = & -iE_{j}(\eta_{i})|E_{j}^{2}(\eta_{i})\rangle2\sqrt{E_{j}(\eta_{i})H(\eta_{i})}a^{3/2}(\eta_{i})\\
 & \approx & -iE_{j}(\eta_{i})\mathcal{N}_{j}\left(\begin{array}{c}
a_{j}\\
b_{j}\\
c_{j}\\
d_{j}
\end{array}\right)\label{eq:derivbc}
\end{eqnarray}
where 
\begin{equation}
d_{j}=\begin{cases}
1 & j\in\{1,2\}\\
\lambda_{22} & j\in\{3,4\}
\end{cases}.
\end{equation}
Note the factor of $1/\sqrt{2}$ in Eq.~(\ref{eq:normalization})
has been taken into account.

To find the WKB solution, it is convenient to eliminate the damping
term in the equation of motion. Define the conformal time $\tau$
as 
\begin{equation}
d\tau a(\tau)\equiv dt=d\eta/H.
\end{equation}
In conformal time, the mode equation becomes 
\begin{equation}
\partial_{\tau}^{2}\Psi^{(j)}+W^{2}\Psi^{(j)}=0
\end{equation}
where 
\begin{equation}
W^{2}\equiv M^{2}(\tau)H^{2}a^{2}-\frac{\partial_{\tau}^{2}a}{a}I
\end{equation}
\begin{equation}
\Psi^{(j)}\equiv a\psi^{(j)}.
\end{equation}
Because $M^{2}$contains $[k^{2}/(aH)^{2}]I$, the eigenvectors of
$W^{2}$ are the same as those of $M^{2}$ with the replacement 
\begin{equation}
k^{2}\rightarrow\bar{k}^{2}\equiv k^{2}-\frac{\partial_{\tau}^{2}a}{a}=k^{2}-\left(\dot{a}^{2}(t)+a(t)\ddot{a}(t)\right)
\end{equation}
and the eigenvalues of $W^{2}$ are 
\begin{equation}
\omega_{j}^{2}(\tau)\equiv\left(Ha\right)^{2}E_{j}^{2}(\tau)\,\,\,\,\,\,\,\mbox{with the replacement }k^{2}\rightarrow k^{2}-2a^{2}H^{2}\mbox{ in }W_{D}^{2}.
\end{equation}

The leading adiabatic order WKB solution is then 
\begin{equation}
\Psi^{(j)(0)}(\tau)=\frac{1}{2}\frac{\mathcal{V}^{(j)}(\tau)}{\sqrt{\omega_{j}(\tau)}}\exp\left[-i\int_{\tau_{i}}^{\tau}d\tau''\omega_{j}(\tau'')\right]\label{eq:WKBsolution}
\end{equation}
where 
\begin{equation}
\mathcal{V}^{(j)}(\tau)=\mathcal{N}_{j}(\tau)\left(\begin{array}{c}
a_{j}(\tau)\\
b_{j}(\tau)\\
c_{j}(\tau)\\
d_{j}(\tau)
\end{array}\right)\label{eq:vdef}
\end{equation}
and the normalization is consistent with Eqs.~(\ref{eq:valbc}) and
(\ref{eq:derivbc}). The solutions $\Psi^{(j)(0)}$ will no longer
be a good approximation when 
\begin{equation}
\left[\frac{1}{2}\frac{\partial_{\tau}^{2}\omega_{j}}{\omega_{j}^{3}}-\frac{3}{4}\left(\frac{\partial_{\tau}\omega_{j}}{\omega_{j}^{2}}\right)^{2}\right]\mathcal{V}^{(j)}-\frac{\partial_{\tau}^{2}\mathcal{V}^{(j)}}{\omega_{j}^{2}}+\frac{\partial_{\tau}\omega_{j}}{\omega_{j}^{3}}\partial_{\tau}\mathcal{V}^{(j)}+\frac{2i}{\omega_{j}}\partial_{\tau}\mathcal{V}^{(j)}\gg\mathcal{V}^{(j)}\label{eq:goodwkb}
\end{equation}
for the nonzero elements of $\mathcal{V}^{(j)}$. Note that when $\tilde{\Phi}_{+}(t)/F_{a}\gg1$,
the WKB solution Eq.~(\ref{eq:WKBsolution}) will be a good approximation
for heavy modes (i.e. modes $E_{j}(k=0)\gg1$).

For small $\omega_{j}$ functions (\emph{i.e.} $j=1,3$), breakdown
of $\Psi^{(j)(0)}$ approximation can occur due to terms such as $(\partial_{\tau}\omega_{j}/\omega_{j}^{2})^{2}$
in Eq.~(\ref{eq:goodwkb}) before the nonadiabaticity associated
with $\partial_{\tau}^{2}\mathcal{V}^{(j)}$ (i.e. turning points
of the usual 1-dimensional WKB approximation). To separate these distinct
nonadiabatic behaviors, we define a semi-numerical mode function $\Psi_{{\rm trial}}^{(j)}(\tau)\equiv\mathcal{V}^{(j)}(\tau)f^{(j)}(\tau)$
where $f^{(j)}(\tau)$ satisfies the scalar mode equation 
\begin{equation}
\partial_{\tau}^{2}f^{(j)}+\omega_{j}^{2}(\tau)f^{(j)}=0\label{eq:modeeq}
\end{equation}
with adiabatic boundary conditions: 
\begin{equation}
f^{(j)}(\tau_{i})=\frac{1}{2\sqrt{\omega_{j}(\tau_{i})}}\,\,\,\,\,\,\,\,\,\,\,\,\,\,\,\,\partial_{\tau}f^{(j)}(\tau_{i})=-i\frac{1}{2}\sqrt{\omega_{j}(\tau_{i})}.
\end{equation}
(In practice, we solve this equation also in the variable $\eta\equiv Ht$
instead of in the conformal time $\tau$.) Note that $\Psi_{{\rm trial}}^{(j)}(\tau)$
satisfies the mode equations whenever $\partial_{\tau}^{2}\mathcal{V}^{(j)}(\tau)$
and $\partial_{\tau}\mathcal{V}^{(j)}(\tau)$ vanishes. Hence, we
can define a measure of how well $\Psi_{{\rm trial}}^{(j)}$ satisfies
the mode equation as 
\begin{equation}
\mathcal{E}_{{\rm na}(j)}(\tau)\equiv\frac{\left|\left[\partial_{\tau}^{2}+W^{2}(\tau)\right]\Psi_{{\rm trial}}^{(j)}\right|}{f^{(j)}\omega_{{\rm eff}}^{2}}
\end{equation}
where $\omega_{{\rm eff}}^{2}$ should be of order $\omega_{j}^{2}$
to measure the validity of the approximation. Since the dominant contribution
to $\mathcal{E}_{{\rm na}(j)}(\tau)$ should come from the the lightest
$\omega_{j}$ which has effective mass terms of order $(c_{+}-2)a^{2}H^{2}$
and since we will be concerned with the $j$ that maximizes $\mathcal{E}_{{\rm na}(j)}(\tau)$,
we choose a positive definite quantity 
\begin{equation}
\omega_{{\rm eff}}^{2}\equiv\sqrt{k^{4}+(c_{+}-2)^{2}(a^{2}H^{2})^{2}}
\end{equation}
independently of $j$. For a computation accurate to about 5\%, we
can define the time $\eta_{NA}$ (in the variable $\eta=Ht$) satisfying
\begin{equation}
\max_{j}\mathcal{E}_{na(j)}(\tau(\eta_{NA}))=O(0.05)\label{eq:errorthreshold}
\end{equation}
after which one can no longer use $\Psi_{{\rm trial}}^{(j)}(\tau)$
for all $j$ as an approximation. For $\eta>\eta_{NA}$, we numerically
solve Eq.~(\ref{eq:originalmodeeq}) with the boundary conditions
\begin{equation}
\psi(\eta_{NA})=\mathcal{V}^{(j)}(\tau(\eta_{NA}))\left[\frac{f^{(j)}(\tau(\eta_{NA}))}{a(\tau(\eta_{NA}))}\right]
\end{equation}
\begin{equation}
\psi'(\eta_{NA})=\partial_{\eta}\left[\mathcal{V}^{(j)}(\tau(\eta))\left[\frac{f^{(j)}(\tau(\eta))}{a(\tau(\eta))}\right]\right]_{\eta=\eta_{NA}}.
\end{equation}

\section{\label{sec:numerical_calculations}Numerical calculations}

\subsection{\label{subsec:homogeneous_solution}Homogeneous solution}

Since the evolution of the background fields $\tilde{\Phi}_{\pm}$
is described by coupled non-linear equations of motion in which $\tilde{\Phi}_{+}$
and $\tilde{\Phi}_{-}$ initially differ by $2\log_{10}(\Mpl/\Fa)\approx15$
orders of magnitude, we integrate the system numerically using the
Class Library for Numbers (CLN) arbitrary-precision arithmetic package~\cite{CLN}.
This is done using our own implementation of a $4$th-order Runge-Kutta-Fehlberg
ordinary differential equation solver with adaptive step size control,
using $50$ digits of precision and a numerical tolerance of $10^{-20}$.

\begin{figure}
\centering{}\includegraphics[width=3.3in]{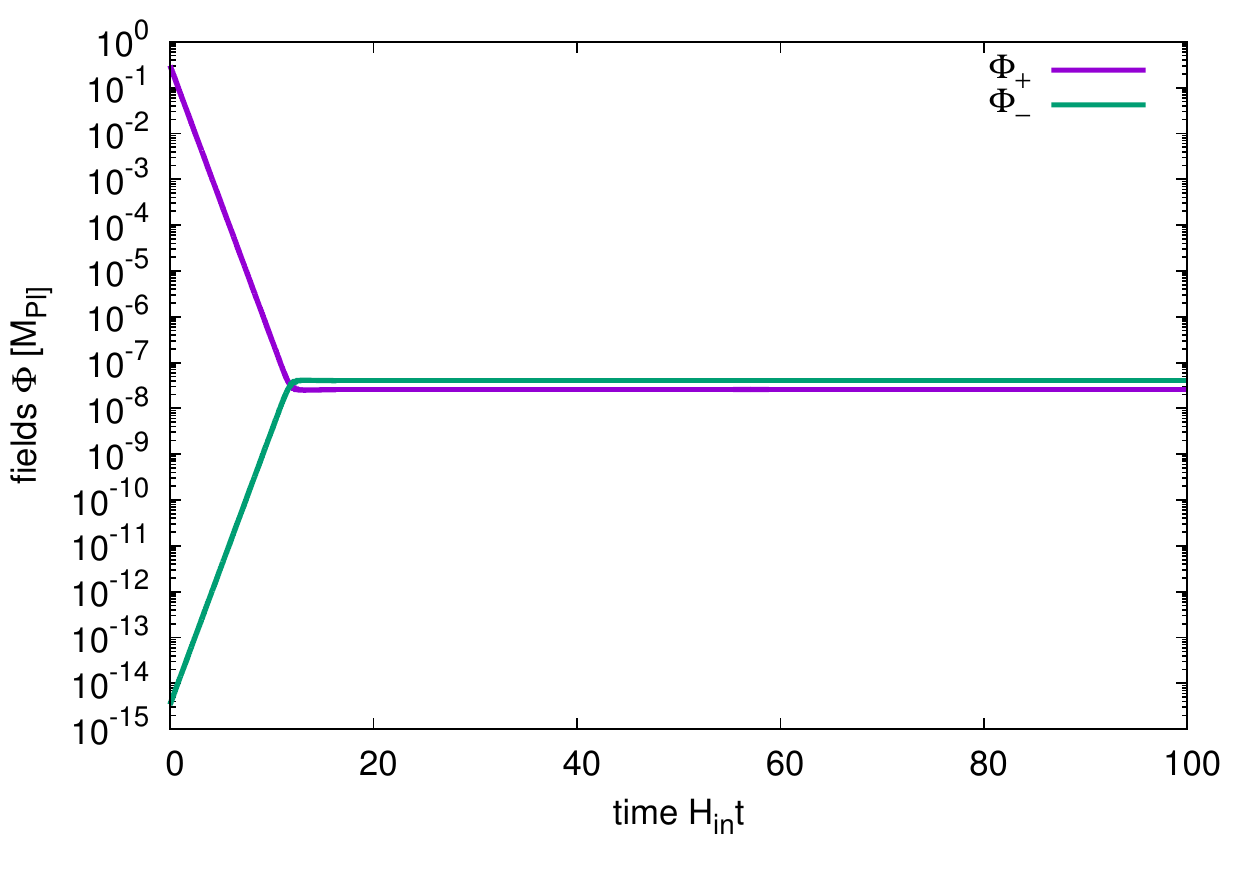}\includegraphics[width=3.3in]{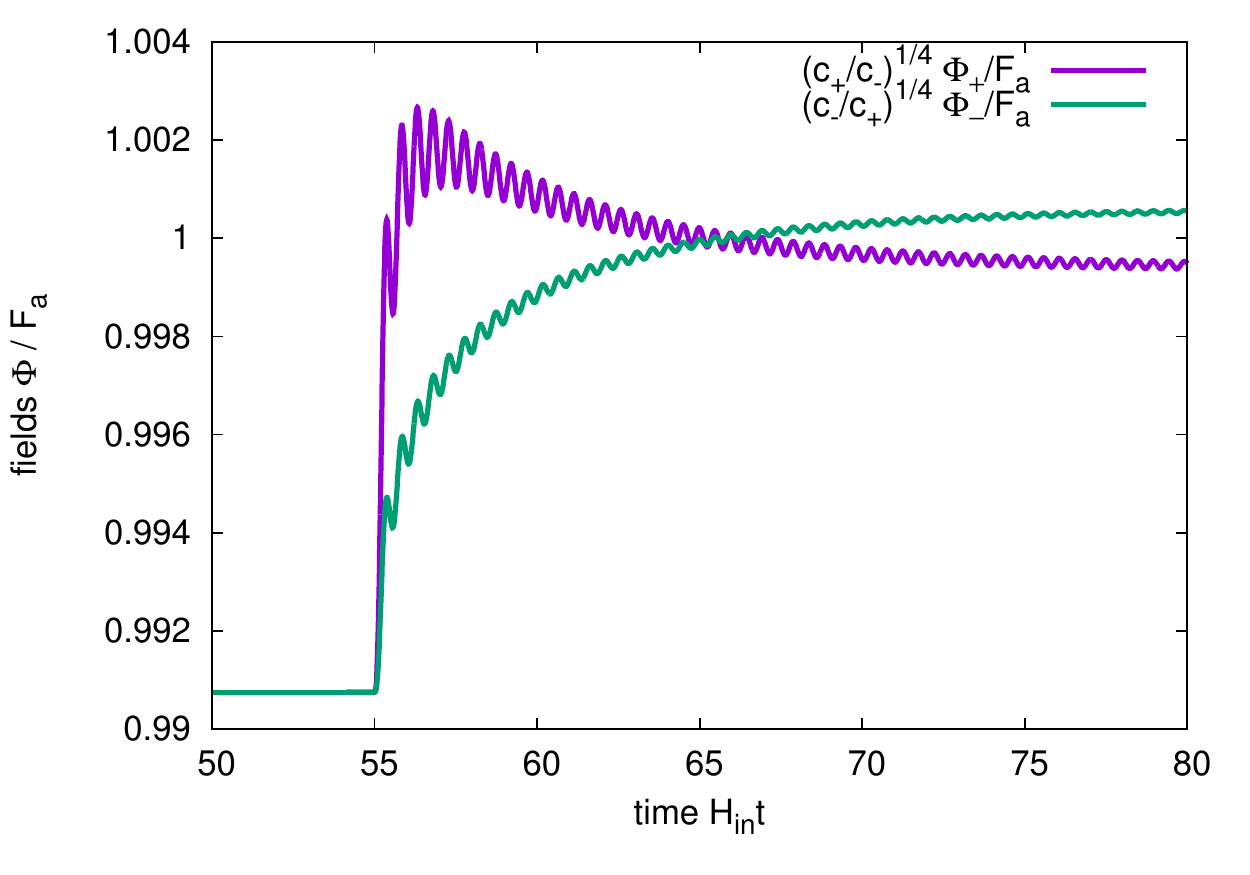}\protect\protect\caption{Homogeneous background field values for a model with $c_{+}=2.235$,
$c_{-}=0.9$, $\thp=0.04$, $h=1$, $\Fa=7.9\times10^{10}$~GeV,
$\Hin=9\times10^{9}$~GeV, and $\Ppin=0.3~\Mpl$. (Left)~The fields
relax to $\sim\Fa$ in $\sim10$ e-folds. (Right)~After inflation
ends at $H_{{\rm in}}t=55$, the fields undergo damped oscillation.
On the horizontal axis of these plots, we have denoted $H_{{\rm in}}$
as the expansion rate at the initial time $t_{i}=0$ to emphasize
the fact that the plots actually continue to the time period after
the end of inflation when the expansion rate starts to decrease. In
the text where we primarily discuss the approximately constant expansion
rate during inflation, we denote $H_{{\rm in}}$ simply as $H$. \label{f:Phi_homogeneous} }
\end{figure}

Figure~\ref{f:Phi_homogeneous} shows our results for a particular
model. Clearly evident are three different regimes of field evolution:
early inflation, late inflation, and post-inflation. During the early
inflationary period, $\Pp\Pm\approx\Fa^{2}$ to excellent precision
as $\Pp$ rolls down its potential. In the approximation of neglecting
inflationary slow-roll parameters, $\Pp''+3\Pp'+\cp\Pp=0$, implying
$\Pp(\eta)=\Ppin\exp(-\gamma\eta)$ for constant $\gamma$. Then $\Pp\Pm\approx\Fa^{2}$
implies $\Pm\propto\exp(\gamma\eta)$. We may approximate $\Pm$ early
during inflation by applying this ansatz to the $\Pm$ equation of
motion, leading to the early-inflation approximations 
\begin{eqnarray}
\gamma & = & \frac{3}{2}\left(1-\sqrt{1-\frac{4}{9}\cp}\right)\label{e:early_inflation_gamma}\\
\Pp(\eta) & = & \Ppin\exp(-\gamma\eta)\label{e:early_inflation_pplus}\\
\Pm(\eta) & = & \frac{\Fa^{2}\Pp}{\Pp^{2}+\left(\gamma^{2}+3\gamma+\cm\right)\Hin^{2}/h^{2}}.\label{e:early_inflation_pminus}
\end{eqnarray}

Once $\tilde{\Phi}_{\pm}\sim\Fa$, the field values stabilize at the
following potential minima: 
\begin{equation}
\frac{\tilde{\Phi}_{\pm}}{\Fa}=\left(\frac{c_{\mp}}{c_{\pm}}\right)^{1/4}\sqrt{1-\frac{\sqrt{\cp\cm}\Hin^{2}}{h^{2}\Fa^{2}}}.\label{e:late_inflation_phi_con}
\end{equation}
In the small-$\Hin/\Fa$ approximation, the time and wavenumber associated
with the transition from early to late inflation can be found by setting
the early-inflation $\Pp$ approximation equal to $\Fa(\cm/\cp)^{1/4}$:
\begin{eqnarray}
\eta_{\star} & = & \frac{1}{\gamma}\log\left[\frac{\Ppin}{\Fa}\left(\frac{\cp}{\cm}\right)^{1/4}\right]\label{e:eta_star}\\
\frac{k_{\star}}{\Hin} & = & \exp(\eta_{\star})=\left(\frac{\Ppin}{\Fa}\right)^{\frac{1}{\gamma}}\left(\frac{\cp}{\cm}\right)^{\frac{1}{4\gamma}}\label{e:k_star}
\end{eqnarray}
where we have set $a(\eta=0)=1$. All $k/H$ in this section can be
interpreted as $k/[Ha(\eta=0)]$.

\begin{figure}
\centering{}\includegraphics[width=3.3in]{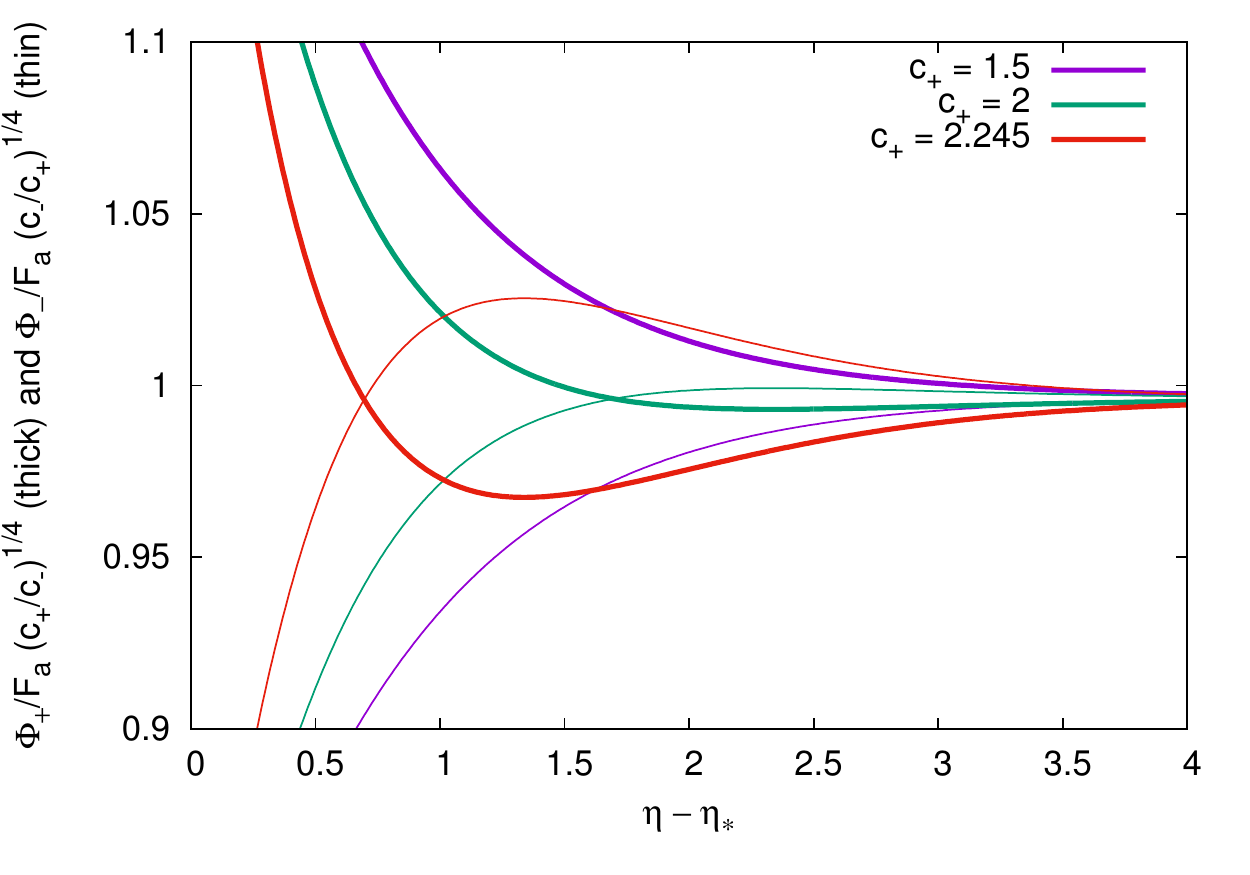}\protect\protect\caption{ The transition from rolling to constant fields is not always monotonic.
$\Pp$ (thick lines) and $\Pm$ (thin lines) overshoot their final
values of Eq.~(\ref{e:late_inflation_phi_con}) for sufficiently
large $\cp$. \label{f:background_bump} }
\end{figure}

The transition between the two regimes is not always monotonic. Figure~\ref{f:background_bump}
shows that for large $\cp$, $\tilde{\Phi}_{\pm}$ overshoot their
final values. We will see that an accurate computation of this overshoot
is necessary for calculating the final power spectrum.

\subsection{\label{subsec:inhomogeneous_solution}Inhomogeneous solution}

\begin{figure}
\centering{}\includegraphics[width=3.3in]{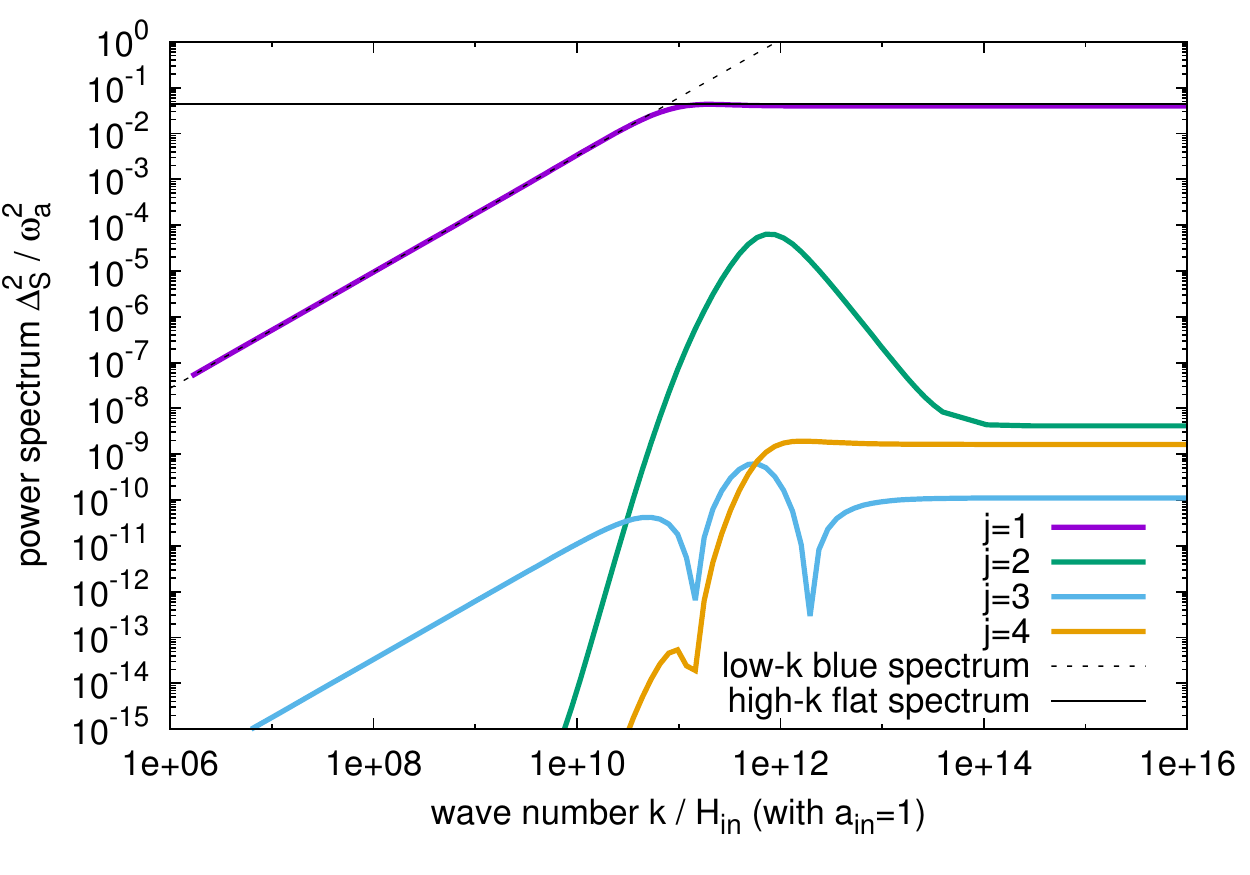}\includegraphics[width=3.3in]{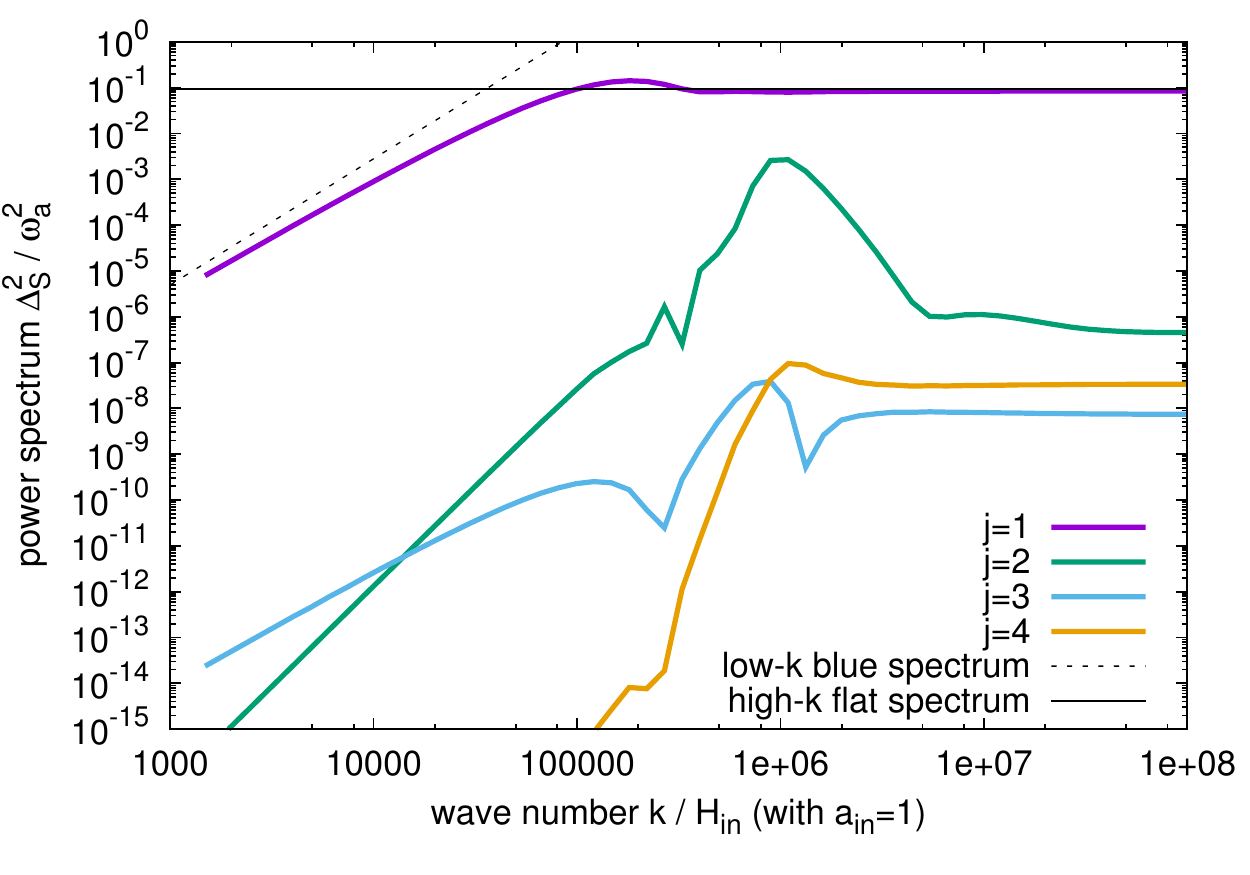}\protect\protect\caption{Power spectra $\Delta_{S}^{2}(k)/\omega_{a}^{2}$ for $\cp=1.5$ and
$\Hin=6\times10^{9}$~GeV (left) as well as $\cp=2.235$ and $\Hin=9\times10^{9}$~GeV
(right), corresponding to spectral indices of $\niso=2.27$ and $\niso=3.76$,
respectively (shown as dotted lines). The other parameters, $h=1$,
$\thp=0.04$, $\cm=0.9$, and $\Fa=7.9\times10^{10}$~GeV are the
same for both models. (See Fig.~\ref{f:Phi_homogeneous} for an explanation
of $H_{{\rm in}}$.) These power spectra are evaluated at $H_{{\rm in}}t=100$,
long after the end of inflation. \label{f:power_spectrum} }
\end{figure}

Given the homogeneous solution for each parameter set, we integrate
the equations for linearized perturbations in $\Phi_{\pm}$ to find
the power spectrum. The set of equations and the procedure used is
described in Sec.~\ref{sec:Setup-of-the}. The initial condition
described in Eqs.~(\ref{eq:valbc}) and (\ref{eq:derivbc}) for each
$k$ mode is set at time $\eta_{i\,{\rm num}}(k)$ when 
\begin{equation}
\Rinit\equiv\frac{k}{a(\eta_{i\,{\rm num}}(k))H}=10.
\end{equation}
Our results are shown in Figure~\ref{f:power_spectrum} for a model
with a soft blue spectrum $\niso=2.27$ and one with a hard blue spectrum
$\niso=3.76$. First, note that the sum of all four modes is dominated
by the $j=1$ mode, which at early times is dominated by $I_{+}$,
and at late times is a mixture of $I_{-}$ and $I_{+}$. Since we
are interested in approximating the power spectrum at the $\approx10\%$
level, we can neglect all but the $j=1$ mode henceforth.

Secondly, the hard blue power spectrum has a peak corresponding to
the transition from blue to flat. Evidently, $\Delta_{S}^{2}(k)/\omega_{a}^{2}$
overshoots its large-$k$ value and then falls back down, with a width
of around an e-fold. This peak is a distinct feature that can facilitate
the detection or exclusion of such models.

\begin{figure}
\centering{}\includegraphics[width=3.3in]{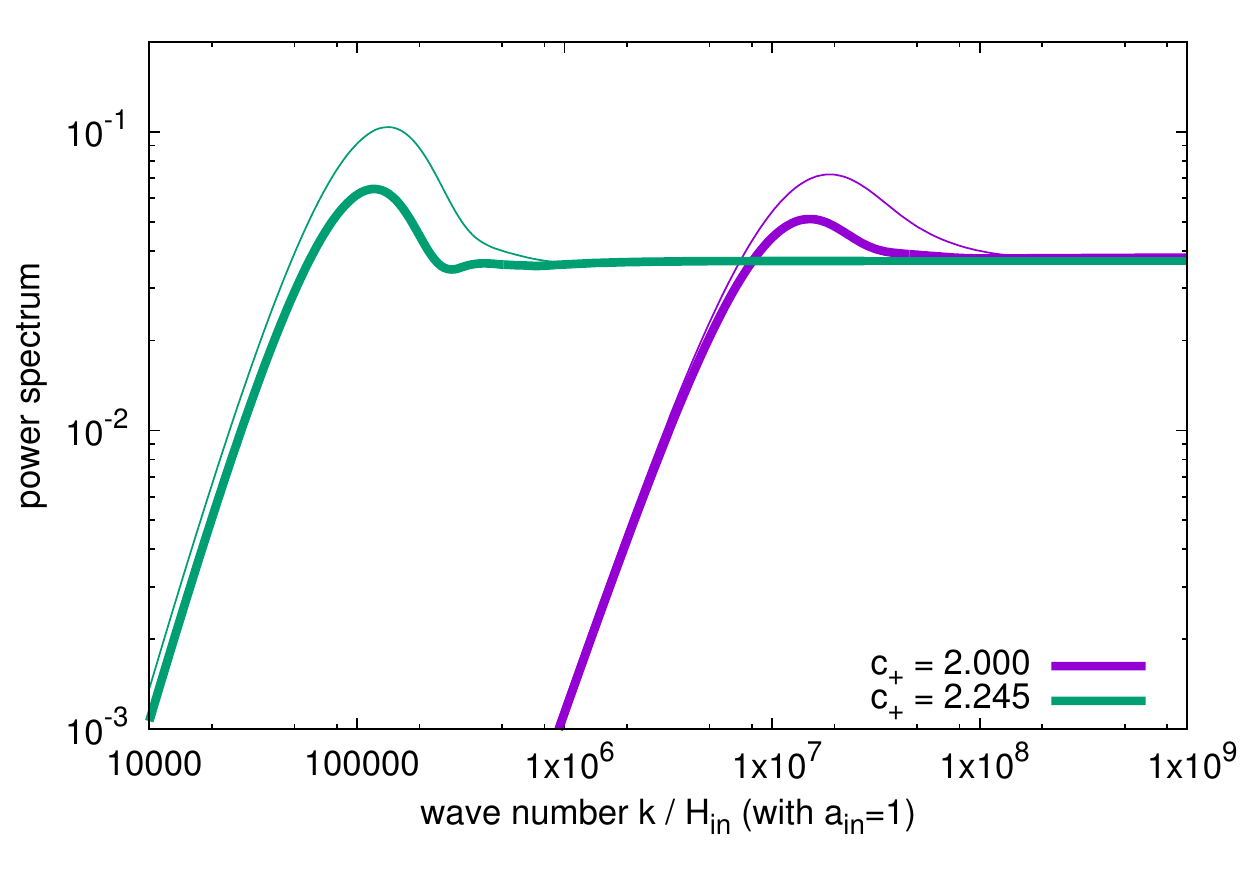}\protect\protect\caption{Peaks of $\Delta_{S}^{2}(k)/\omega_{a}^{2}$ , calculated using the
numerical solutions of Sec.~\ref{subsec:homogeneous_solution} (thick
lines) as well as a smoothed homogeneous solution interpolating Eqs.~(\ref{e:early_inflation_pplus}),
(\ref{e:early_inflation_pminus}), and (\ref{e:late_inflation_phi_con})
(thin lines). (See Fig.~\ref{f:Phi_homogeneous} for an explanation
of $H_{{\rm in}}$.) These plots are evaluated at $H_{{\rm in}}t=100$,
long after the end of inflation. \label{f:peak_num_vs_smooth} }
\end{figure}

In order to investigate the dependence of this power spectrum peak
on the transition from rolling to constant homogeneous fields $\Phi_{\pm}$,
we constructed a smoothed approximation to the numerically-computed
homogeneous fields of Sec.~\ref{subsec:homogeneous_solution}. We
approximated $\Pp$ by adding Eqs.~(\ref{e:early_inflation_pplus})
and (\ref{e:late_inflation_phi_con}) in quadrature, and $\Pm$ by
replacing $\gamma$ by $-d\log(\Pp)/d\eta$ in Eq.~(\ref{e:early_inflation_pminus}).
Figure~\ref{f:peak_num_vs_smooth} compares the power spectra resulting
from the numerically-computed fields to their smoothed counterparts.
Evidently the smoothing enhances the amplitude of the power spectrum
feature by a factor of about two, even though the overshoot feature
seen in Fig.~\ref{f:background_bump} is only a $5\%-10\%$ effect.
Thus the accurate numerical computations of the homogeneous background
fields in the nonadiabatic region are necessary even for $\approx10\%$
accuracy in the final power spectrum.

\begin{figure}
\centering{}\includegraphics[width=3.3in]{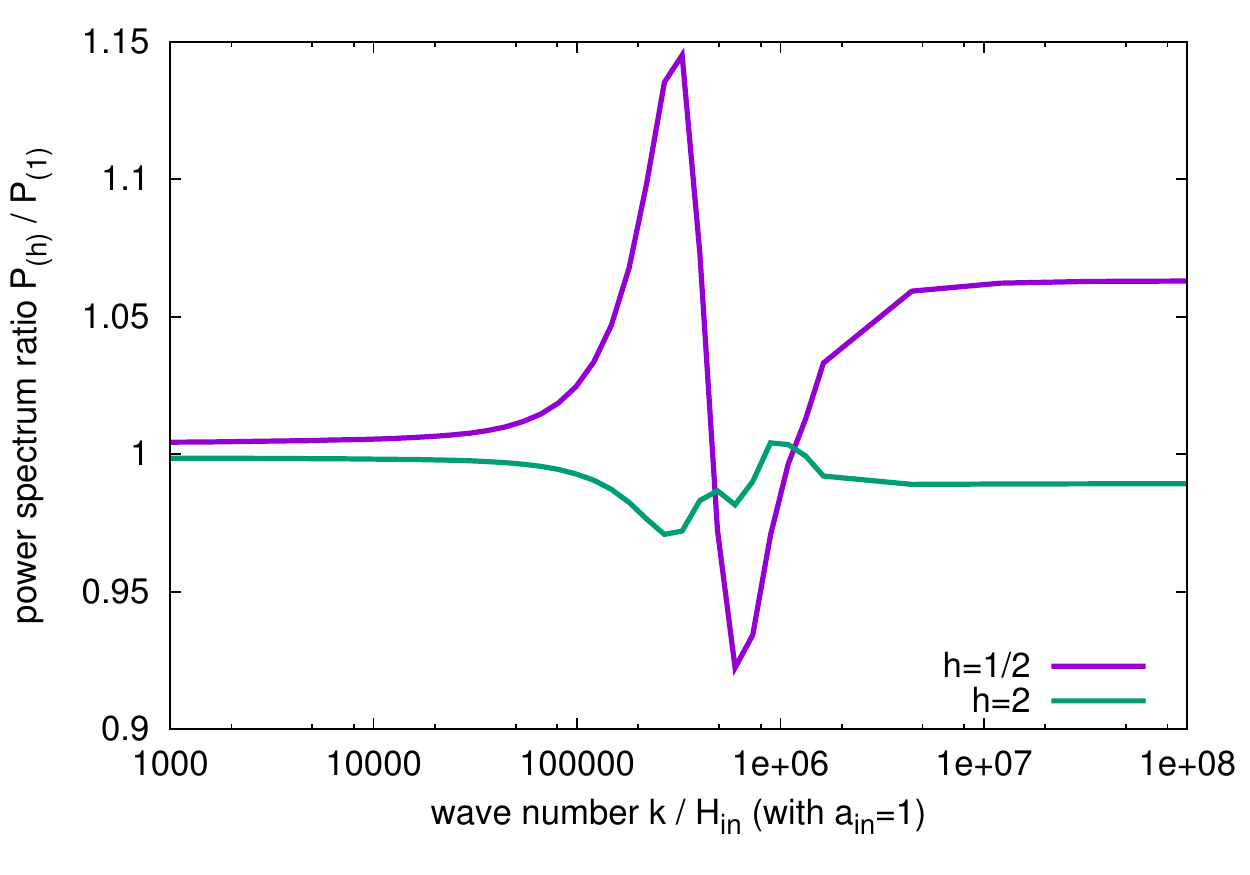}\protect\protect\caption{ Fractional change in the power spectrum due to the variation of $h$
from $1$. The remaining parameters are identical to those used in
Figure~\ref{f:power_spectrum}~(bottom). (See Fig.~\ref{f:Phi_homogeneous}
for an explanation of $H_{{\rm in}}$.) These plots are evaluated
at $H_{{\rm in}}t=100$, long after the end of inflation.\label{f:vary_h} }
\end{figure}

Figure~\ref{f:vary_h} shows that factor-of-two changes in $h$ result
in only small changes to the power spectrum. Decreasing or increasing
$h$ has the effect of shifting the power spectrum peak slightly to
the left or the right, respectively. Since we have no compelling reason
for choosing $h$ an order of magnitude away from unity, and we are
interested in a power spectrum calculation at the $\approx10\%$ level,
we do not study $h$ further.

\begin{figure}
\centering{}\includegraphics[width=3.3in]{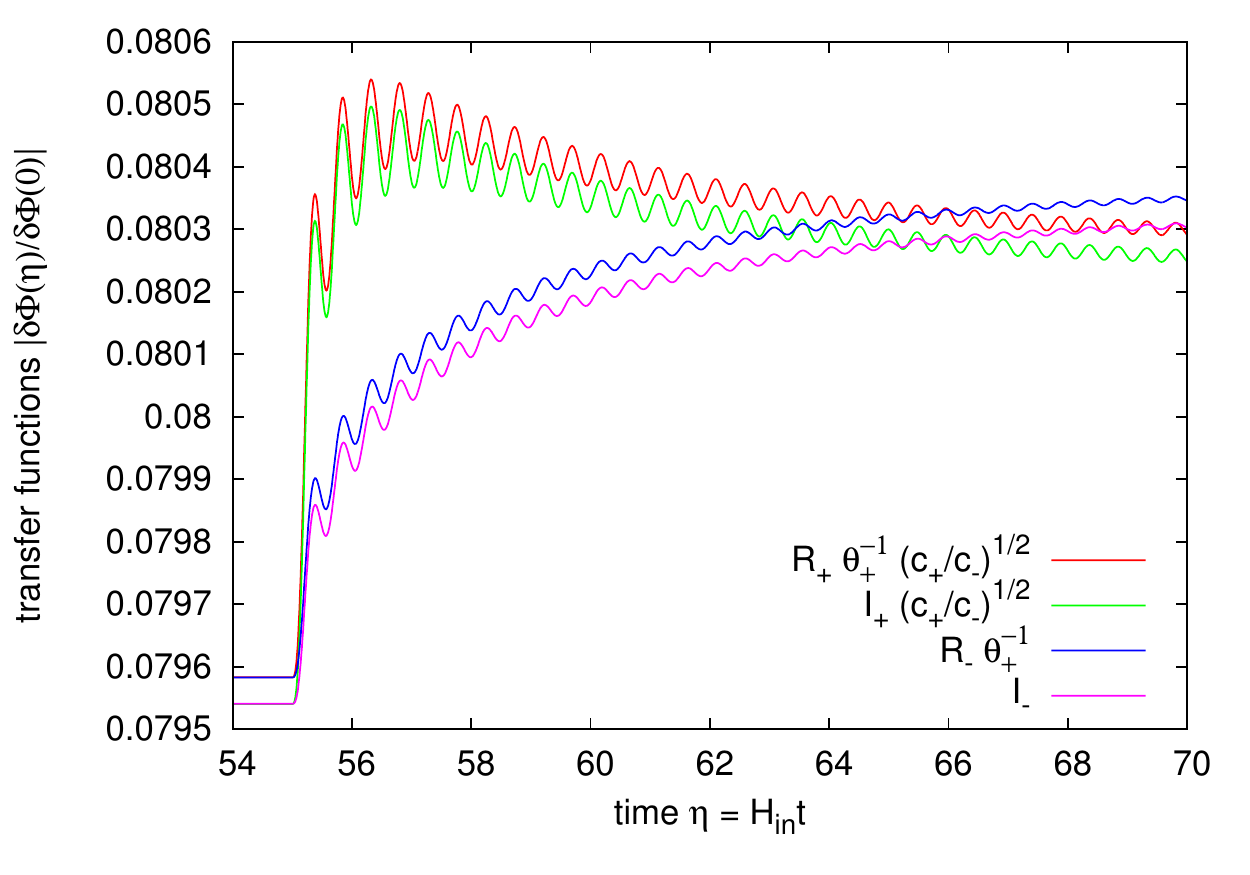}\protect\protect\caption{ Field perturbations for a single mode after the end of inflation,
$\eta=55$, for the $\cp=2.235$ model shown in Fig.~\ref{f:Phi_homogeneous}.
(See Fig.~\ref{f:Phi_homogeneous} for an explanation of $H_{{\rm in}}$.)\label{f:inflation_end} }
\end{figure}

Finally, we investigate the evolution of the perturbations through
the end of inflation. Figure~\ref{f:inflation_end} shows that they
increase in magnitude by $\approx\Hin^{2}/\Fa^{2}\approx1\%$ after
the end of inflation, and undergo damped oscillations at the $\approx0.1\%$
level, similar to the homogeneous fields $\Phi_{\pm}$ in Fig.~\ref{f:Phi_homogeneous}.
Thus the perturbations evolve adiabatically through the end of inflation,
without any significant suppression or amplification due to oscillations
in $\Phi_{\pm}$.

\subsection{\label{subsec:fitting_function}Fitting function}

As we saw in the previous subsection, the power spectrum is sensitively
dependent on the details of the homogeneous field evolution, which
is computationally expensive to determine. Furthermore, we would like
to constrain blue-tilted isocurvature models such as this one using
large-scale structure data. A power spectrum needing several parameters
to describe its broad features and several more for the peak feature
would be poorly constrained by the data.

In this section we construct a fitting function describing the power
spectrum in terms of 3 independent parameters: the blue tilt $\niso=2\gamma+1$,
determined by $\cp$; the transition scale $k_{\star}/\Hin$, given
in Eq.~(\ref{e:k_star}); and the overall amplitude, as we explain
further below. We fit the other features of the power spectrum, including
the amplitude and the width of the peak associated with the blue-to-flat
transition, in terms of these three. The scaling of our parameters
with $\cm$ is also given. Since varying $h$ by a factor of two is
found to result in $<10\%$ changes to the power spectrum, we fix
$h=1$ henceforth.

As a starting point, we choose a broken power law model characterized
by a high-$k$ amplitude $A$, a blue-to-flat ratio $\rho$, and a
dimensionless width parameter $w$, as $A\left[1+(\rho(k/k_{\star})^{2\gamma})^{-1/w}\right]^{-w}$.
Here $w$ is of the order of the number of e-folds over which the
transition from blue to flat takes place. Judging from Fig.~\ref{f:power_spectrum},
$w$ is of order unity, and is larger for bluer tilts. In order to
fit the power spectrum peak, we choose a Lorentzian function characterized
by an amplitude $\alpha$, a center $\mu$, and a width $\sigma$,
as well as a skew parameter $\lambda$ characterizing the asymmetry
of the peak. Our fitting function, valid in the $\thp\ll1$ and $\Hin/\Fa\ll1$
limits, takes the form 
\begin{eqnarray}
\frac{\Pfit(k)}{\omega_{\mathrm{a}}^{2}} & = & 
\frac{\Rinit^{3}\Hin^{2}\sqrt{\cm/\cp}}{4\pi^{2}\Fa^{2}\thp^{2}\sqrt{\Rinit^{2}+\cp-2}}\left|T_{I_{-}}\right|^{2}\label{e:fit_P}\\
\left|T_{I_{-}}\right|^{2} & = & A\frac{1+\alpha L\!\left(\frac{\ln(k/k_{\star})-\mu}{\sigma}\right)S\!\left(\lambda\frac{\ln(k/k_{\star})-\mu}{\sigma}\right)}{\left[1+(\rho(k/k_{\star})^{2\gamma})^{-1/w}\right]^{w}}\label{e:fit_Im}\\
L(x) & = & 1/(1+x^{2})\label{e:fit_L}\\
S(x) & = & 1+\tanh(x).\label{e:fit_S}
\end{eqnarray}

\begin{table}
\begin{footnotesize} \tabcolsep=0.1cm %
\begin{tabular}{r|ccccccc}
$\cp$  & ${\tilde{A}}$  & ${\tilde{\rho}}$  & $w$  & $\alpha$  & $\mu$  & $\sigma$  & $\lambda$ \tabularnewline
\hline 
1.500  & 0.9036  & 1.123  & 0.3558  & 0.1333  & 0.4554  & 0.2894  & 0.01114\tabularnewline
1.600  & 0.9015  & 1.145  & 0.3799  & 0.1743  & 0.3805  & 0.3393  & 0.01116\tabularnewline
1.700  & 0.8989  & 1.168  & 0.425  & 0.2261  & 0.3131  & 0.3779  & 0.008246\tabularnewline
1.800  & 0.896  & 1.188  & 0.4986  & 0.2942  & 0.2518  & 0.4052  & 0.004133\tabularnewline
1.900  & 0.8924  & 1.204  & 0.6192  & 0.3918  & 0.192  & 0.4268  & -0.003754\tabularnewline
1.950  & 0.8904  & 1.208  & 0.7109  & 0.4614  & 0.1606  & 0.4398  & -0.01367\tabularnewline
2.000  & 0.8885  & 1.208  & 0.8426  & 0.5595  & 0.1264  & 0.4601  & -0.03494\tabularnewline
2.025  & 0.8876  & 1.205  & 0.9328  & 0.628  & 0.1078  & 0.4756  & -0.05327\tabularnewline
2.050  & 0.8873  & 1.198  & 1.054  & 0.723  & 0.08889  & 0.4994  & -0.08532\tabularnewline
2.075  & 0.8891  & 1.182  & 1.252  & 0.8889  & 0.07784  & 0.5481  & -0.1664\tabularnewline
2.100  & 0.8926  & 1.158  & 1.574  & 1.212  & 0.06334  & 0.6359  & -0.2712\tabularnewline
2.125  & 0.8968  & 1.125  & 1.965  & 1.717  & 0.02296  & 0.7236  & -0.306\tabularnewline
2.150  & 0.9019  & 1.086  & 2.338  & 2.304  & -0.01774  & 0.777  & -0.2973\tabularnewline
2.175  & 0.9077  & 1.036  & 2.715  & 2.985  & -0.05606  & 0.8078  & -0.2734\tabularnewline
2.200  & 0.9148  & 0.9593  & 3.173  & 3.934  & -0.1002  & 0.8302  & -0.2392\tabularnewline
2.210  & 0.9185  & 0.9129  & 3.401  & 4.46  & -0.1208  & 0.838  & -0.2228\tabularnewline
2.220  & 0.9229  & 0.8481  & 3.672  & 5.129  & -0.1435  & 0.8452  & -0.2049\tabularnewline
2.230  & 0.9284  & 0.7483  & 4.017  & 6.063  & -0.1699  & 0.8528  & -0.1845\tabularnewline
2.235  & 0.932  & 0.6735  & 4.236  & 6.699  & -0.1852  & 0.8565  & -0.173\tabularnewline
2.240  & 0.9365  & 0.567  & 4.509  & 7.553  & -0.2028  & 0.8604  & -0.1599\tabularnewline
2.245  & 0.9429  & 0.3962  & 4.895  & 8.877  & -0.2254  & 0.8651  & -0.1438\tabularnewline
2.249  & 0.9525  & 0.1323  & 5.481  & 11.2  & -0.2556  & 0.8709  & -0.1237\tabularnewline
\end{tabular}\end{footnotesize} \protect\protect\caption{ Best-fitting parameters for the fitting function of Eqs.~(\ref{e:fit_P}),
(\ref{e:fit_Im}), (\ref{e:fit_L}), and (\ref{e:fit_S}), with $A$
and $\rho$ expressed in terms of the rescaled amplitudes ${\tilde{A}}=A\Rinit^{3}(1+\cm/\cp)/\sqrt{\Rinit^{2}+\cp-2}$
and ${\tilde{\rho}}=2\pi\rho/[2^{2\nu}\Gamma(\nu)^{2}(1+\cp/\cm)]$
with $\nu=\sqrt{9/4-\cp}$. \label{t:best-fit_parameters} }
\end{table}

The seven parameters in $\left|T_{I_{-}}\right|^{2}$ are determined
by minimizing the mean-squared difference between $\Pfit(k)$ and
numerical computations over $100$ logarithmically-spaced bins in
the range $10^{-3}k_{\star}\leq k\leq10^{3}k_{\star}$. Table~\ref{t:best-fit_parameters}
shows our results for a range of $\cp$ values from $1.5$ to $2.249$,
corresponding to $2.27\leq\niso\leq3.94$. Note that $\niso\gtrsim2.4$
is phenomenologically significant because such blue isocurvature spectral
indices require a time dependent mass transition \cite{Chung:2015tha}
for the isocurvature field degree of freedom just as in the particular
axion model being studied here. Table~\ref{t:best-fit_parameters}
can be used for computing the isocurvature spectrum for a continuous
family of model parameters by interpolating the seven parameters $A$,
$\rho$, $w$, $\alpha$, $\mu$, $\sigma$, and $\lambda$ from the
table, finding $\left|T_{I_{-}}\right|^{2}$ using Eqs.~(\ref{e:fit_Im}),
(\ref{e:fit_L}), and (\ref{e:fit_S}), and then substituting this
into Eq.~(\ref{e:fit_P}).

\begin{table}
\begin{footnotesize} \tabcolsep=0.1cm %
\begin{tabular}{r|ccc}
Model:  & Small-$\cp$  & Medium-$\cp$  & High-$\cp$ \tabularnewline
\hline 
$\cp$  & 0.1596  & 1.942  & 2.236 \tabularnewline
$\cm$  & 0.597  & 0.855  & 0.548 \tabularnewline
$\thp$  & 0.0877  & 0.0272  & 0.0209 \tabularnewline
$\Fa$~{[}GeV{]}  & \sci{9.30}{11}  & \sci{3.44}{10}  & \sci{1.46}{10}\tabularnewline
$\Hin$~{[}GeV{]}  & \sci{6.57}{10}  & \sci{2.76}{9}  & \sci{7.23}{8} \tabularnewline
$\Ppin/\Mpl$  & 0.226  & 0.452  & 0.117 \tabularnewline
\end{tabular}\end{footnotesize} \protect\protect\caption{ Models with randomly-chosen parameters, used for testing the fitting
function of Eqs.~(\ref{e:fit_P}), (\ref{e:fit_Im}), (\ref{e:fit_L}),
and (\ref{e:fit_S}). In all cases, $h=1$ is fixed. \label{t:test_fit}
The dark matter fractions $\omega_{a}$ corresponding to these parametric
choices are $0.027$ (small-$c_{+}$), $5\times10^{-5}$ (medium-$c_{+}$),
and $10^{-5}$ (high-$c_{+}$).}
\end{table}

\begin{figure}
\centering{}\includegraphics[width=3.3in]{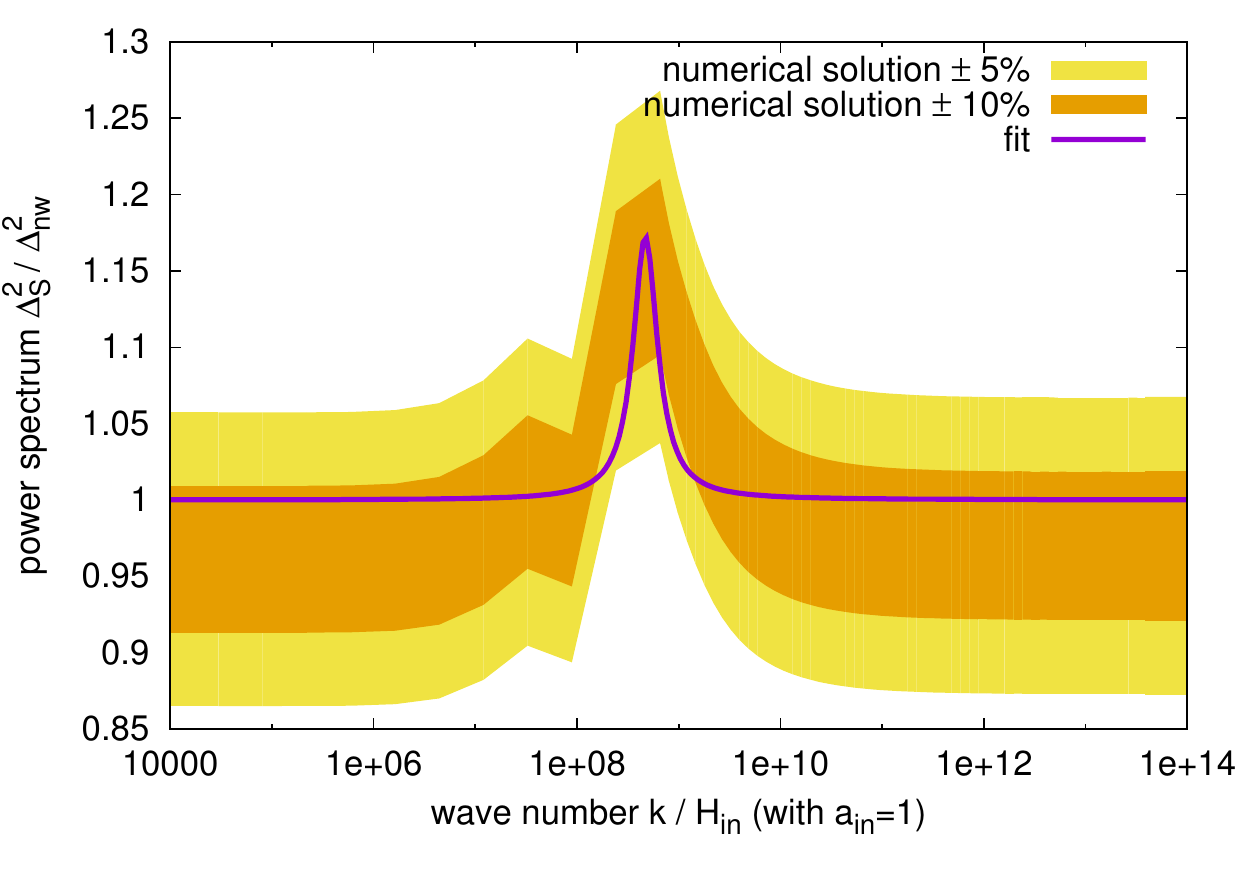}\includegraphics[width=3.3in]{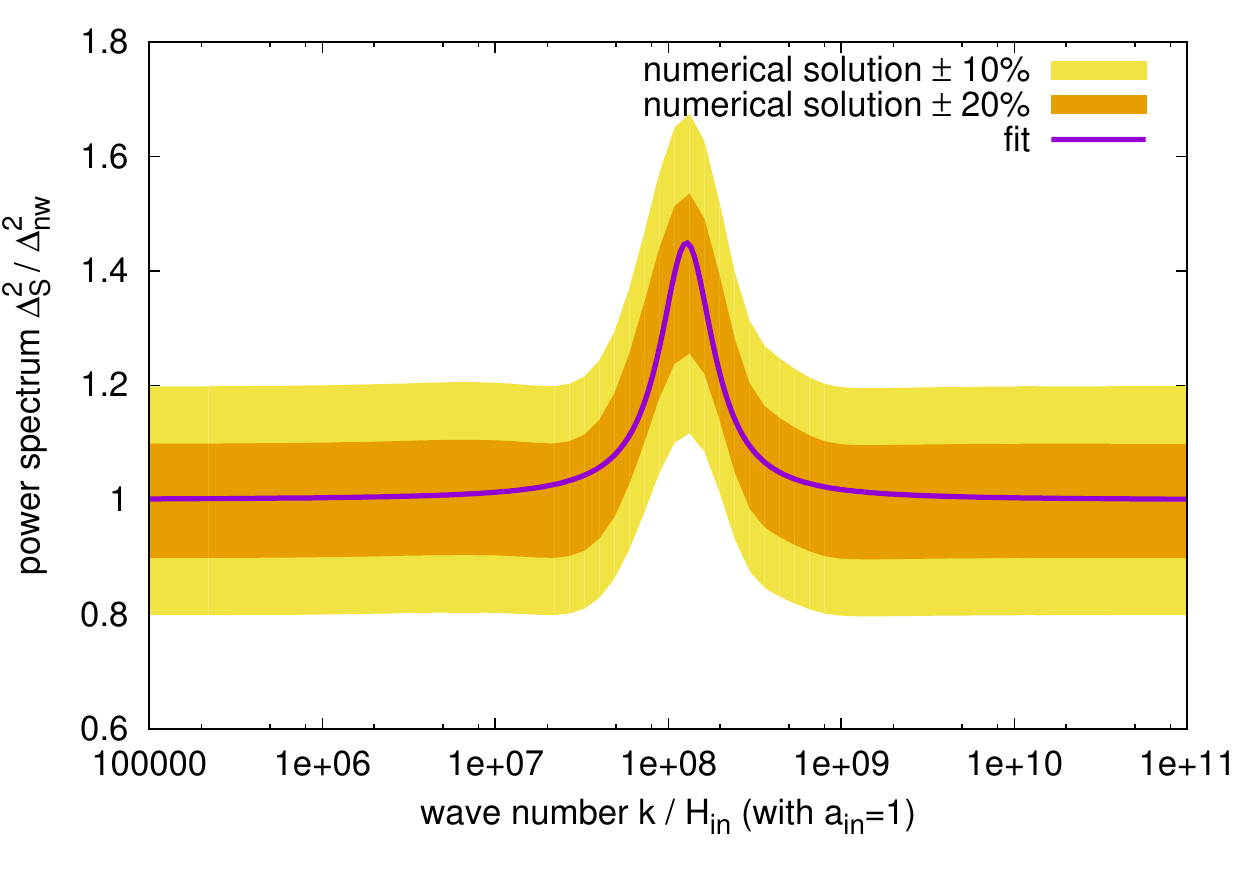}\\
 \includegraphics[width=3.3in]{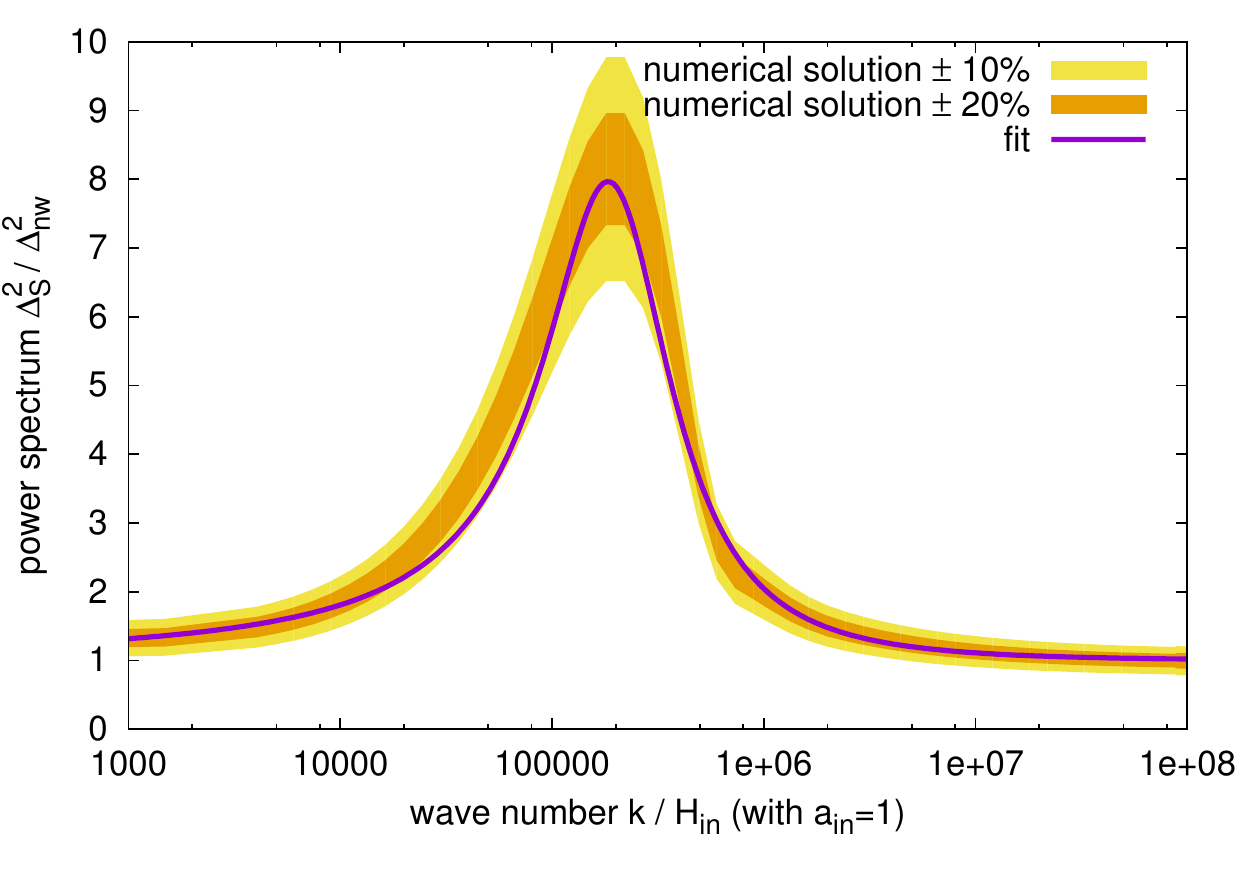}\protect\protect\caption{ Tests of our fitting function for randomly-chosen model parameters,
as given in Table~\ref{t:test_fit}. For clarity, power spectra have
been divided by ``no-wiggle'' spectra $\Delta_{\mathrm{nw}}^{2}$
found by setting $\alpha=0$ in the fitting function. (Top-left)~Small-$\cp$.
(Top-right)~Medium-$\cp$. (Bottom)~High-$\cp$. \label{f:test_fit} }
\end{figure}

Finally, we test our fitting function for three models randomly chosen
to have small ($1.5<\cp<1.7$), medium ($1.9<\cp<2.1$), and large
($2.235<\cp<2.245$) values of $\cp$. The other parameter have been
chosen using uniform random distributions with $0.5<\cm<1$, $0.01<\thp<0.1$,
$10<\log_{10}\Fa\textrm{[GeV]}<12$, $0.01<\Hin/\Fa<0.1$, and $0.1<\Ppin/\Mpl<1$.
The models chosen are listed in Table~\ref{t:test_fit}. Figure~\ref{f:test_fit}
shows the numerically computed power spectra along with our fitting
functions. In each case, the power spectra have been divided by a
``no-wiggle'' power spectrum defined by setting $\alpha=0$ in the
corresponding fitting function. The figure shows that the fitting
function of Eqs.~(\ref{e:fit_P}), (\ref{e:fit_Im}), (\ref{e:fit_L}),
and (\ref{e:fit_S}) is accurate at the $10\%-20\%$ level for a wide
range of models.

Instead of dealing with the axion model parameters directly, it may
be a bit clearer for phenomenology to explicitly choose 3 parameters
for the fits. Note that our fitting function Eq.~(\ref{e:fit_P})
is of the form 
\begin{equation}
F_{{\rm fit\,{\rm gen}}}^{2}(k,k_{\star},n_{I},\mathcal{Q}_{1},\mathcal{Q}_{2})=\mathcal{Q}_{1}\frac{1+\alpha(n_{I})L\left[\frac{1}{\sigma(n_{I})}\ln\left(e^{-\mu(n_{I})}\frac{k}{k_{\star}}\right)\right]S\left[\frac{\lambda(n_{I})}{\sigma(n_{I})}\ln\left(e^{-\mu(n_{I})}\frac{k}{k_{\star}}\right)\right]}{\left[1+\left(\mathcal{Q}_{2}\left(\frac{k}{k_{\star}}\right)^{n_{I}-1}\right)^{-1/w}\right]^{w}}
\end{equation}
where one can make parameters such as $\alpha$ functions of a generic
spectral index parameter $n_{I}$ through the Table \ref{t:best-fit_parameters}
using the map 
\begin{equation}
c_{+}(n_{I})=\frac{1}{4}(n_{I}-1)(7-n_{I}).
\end{equation}
This is naively a function of 4 parameters: $\mathcal{Q}_{1,2}$,
$k_{*}$, and $n_{I}$. However, within the limited range of $c_{-}$
considered here, $\mathcal{Q}_{2}$ is independent of $c_{-}$ at
the level of accuracy we were aiming for. This means $\mathcal{Q}_{2}$
can be extracted from $\tilde{\rho}$ in Table \ref{t:best-fit_parameters}
after fixing $c_{-}=0.9$ used in making the table: \emph{i.e.} 
\begin{equation}
\mathcal{Q}_{2}=\rho(n_{I})=\tilde{\rho}(n_{I})2^{2\sqrt{\frac{9}{4}-c_{+}(n_{I})}}\frac{\Gamma^{2}\left(\sqrt{\frac{9}{4}-c_{+}(n_{I})}\right)}{2\pi}\left(1+\frac{c_{+}(n_{I})}{0.9}\right).
\end{equation}
Hence in hunting for this lamp post model signatures in future data,
we advocate using 
\begin{equation}
\boxed{\Delta_{S}^{2}(k;k_{\star},n_{I},\mathcal{Q}_{1})=F_{{\rm fit\,{\rm gen}}}^{2}\left(\frac{k}{k_{\star}},n_{I},\mathcal{Q}_{1},\rho(n_{I})\right)}\label{eq:lessmodeldep}
\end{equation}
which is explicitly a function of 3 parameters $k_{\star}$, $n_{I}$,
and $\mathcal{Q}_{1}$. It is interesting that even though one would
generically expect that there are at least 5 parameters describing
a break spectrum with a bump (\emph{e.g.} overall amplitude, break
location, flat spectrum amplitude, bump height, and bump width), the
underlying axion model has approximately reduced this to only 3 independent
parameters. After doing such a fit, the interpretation of best fit
$\mathcal{Q}_{1}$ in the context of our axion model would be 
\begin{equation}
\mathcal{Q}_{1}=\left(\frac{\Hin}{2\pi}\right)^{2}\frac{\tilde{A}(c_{+})\sqrt{\cm/\cp}}{\Fa^{2}\thp^{2}(1+\cm/\cp)}\omega_{\mathrm{a}}^{2}
\end{equation}
where $\omega_{a}$ is the dark matter fraction in axions defined
in Eq.~(\ref{eq:darkmatterfraction}) and is approximately 
\begin{equation}
\omega_{\mathrm{a}}\approx W_{a}\thp^{2}\left(\frac{\sqrt{2}F_{a}\sqrt{\frac{c_{-}+c_{+}}{\sqrt{c_{-}c_{+}}}}}{10^{12}{\rm GeV}}\right)^{n_{PT}}
\end{equation}
where we have assumed that $c_{\pm}>0$.

Despite there being an upper limit on the spectral break location
$k_{\star}$ within this axion model coming from the the fact $\Ppin$
is sub-Planckian at initial times, it is not extremely constraining.
For example, setting $\Ppin\lesssim M_{p}$ at initial times, the
constraint on $k_{\star}$ is 
\begin{equation}
\frac{k_{\star}}{H}\lesssim\left(\frac{M_{p}}{\Fa}\right)^{\frac{2}{n_{I}-1}}\left(\frac{\cp}{\cm}\right)^{\frac{1}{2(n_{I}-1)}}.
\end{equation}
If $F_{a}$ is expressed in terms of dark matter fraction $\omega_{a}$
with $\cp=\cm=1$ and $n_{I}=3.94$, this bound becomes 
\begin{equation}
\frac{k_{\star}}{H}\lesssim10^{5}\left(\frac{\thp}{4\times10^{-2}}\right)^{1.14}\left(\frac{\omega_{a}(\thp{\rm \,\, fixed})}{10^{-5}}\right)^{-0.57}.\label{eq:kstarexample}
\end{equation}
The left hand side of this inequality can be interpreted in terms
of length scales today as
\begin{equation}
\frac{k_{\star}}{a({\rm today})}\approx1\,{\rm Mpc}^{-1}e^{-\left(N_{e}-54\right)}\left(\frac{k_{\star}/H}{10^{5}}\right)\left(\frac{T_{{\rm rh}}}{10^{7}{\rm GeV}}\right)^{1/3}\left(\frac{H}{7\times10^{8}{\rm GeV}}\right)^{1/3}\left(\frac{g_{*S}(t_{0})}{3.9}\right)^{1/3}\label{eq:lengthscaletoday}
\end{equation}
where $N_{e}$ is the number of e-folds between $t_{i}$ and the end
of inflation, $T_{{\rm rh}}$ is the reheating temperature, $g_{*S}(t_{0})$
is the effective number of entropy degrees of freedom today.%
\footnote{The main approximation in this formula is the neglect of the slow-roll
evolution of the expansion rate $H$ during inflation. We have also
assumed that there is exactly one inflationary period and its attendant
reheating since the time $t_{i}$.%
} Since the scales accessible to cosmological experiments are approximately
\begin{equation}
10^{-3}\,{\rm Mpc}^{-1}\lesssim k/a({\rm today})\lesssim{\rm Mpc}^{-1},
\end{equation}
we see from Eqs.~(\ref{eq:kstarexample}) and (\ref{eq:lengthscaletoday})
that it is not difficult to arrange the break in a region that is
observable by changing the inflationary/reheating model (\emph{e.g.~}decrease
$T_{{\rm rh}}$ and/or increase the number of e-folds), increasing
the dark matter fraction $\omega_{a}$, and/or decrease the initial
condition field value of $\Ppin$. It is also not difficult to push
$k_{\star}$ outside of the observable region by increasing $\thp$
even with the inflationary/reheating model fixed to the canonical
values shown in Eq.~(\ref{eq:lengthscaletoday}). The extraction
of $\mathcal{Q}_{1}$ will contain information about $\Hin/\Fa$ that
can be varied independently of $\omega_{a}$. Its implication for
the tensor-to-scalar ratio was already explored in \cite{Chung:2015pga}.
Perturbativity and/or the linear fluctuation approximation must be
reanalyzed for $k\gtrsim k_{\star}$ modes when $\Hin/(2\pi\thp\Fa)$
implied by the best fit parameters is larger than unity. Finally,
as noted above, the spectral index $n_{I}\leq3.94$ bound should be
enforced when fitting since only that range has been tabulated Table
\ref{t:best-fit_parameters}.

Given the general field theory model degeneracies that exist if one
fixes only the two-point function, Eq.~(\ref{eq:lessmodeldep}) is
likely to be more general than the specific underlying model used
to inspire it. The utility of this paper is to show that this parameterization
is consistent with at least one realistic underlying field theory
model.

\section{\label{sec:Conclusions}Conclusions}

In observable blue isocurvature spectral models, there is a break
in the spectrum corresponding to the mass of the isocurvature field
undergoing a transition \cite{Kasuya:2009up,Chung:2015tha}. Analytic
techniques typically break down in this spectral regime \cite{Chung:2015pga}
because of a combination of nonadiabatic time-dependence of the mass
matrix and the non-linearity of the time dependent background field
equations which govern the mass matrix. We have computed the isocurvature
perturbations and the observable spectrum for the axion model of \cite{Kasuya:2009up}
numerically in this region. We find a bump near the break in the spectrum
that enhances the blue isocurvature signal by almost a factor of two
for a steep spectral index. We constructed an economical 3-parameter
fitting function Eq.~(\ref{eq:lessmodeldep}) which reproduces the
bump at the 20\% accuracy level. Although this fitting function has
been checked only against the particular axion model studied in this
paper, the qualitative form of the bump connecting two spectral regions
may be generic. Hence, this ``lamp-post'' model computation is likely
to be useful for future hunt for blue isocurvature contributions to
the cosmic microwave background and large scale structure power spectra. 
\begin{acknowledgments}
This work was supported in part by the DOE through grant DE-FG02-95ER40896. 
\end{acknowledgments}
\bibliographystyle{JHEP}
\bibliography{blue_isocurvature,misc,ref,Inflation_general2}

\providecommand{\href}[2]{#2}\begingroup\raggedright\begin{thebibliography}{10}

\bibitem{Copeland:1994vg}
E.~J. Copeland, A.~R. Liddle, D.~H. Lyth, E.~D. Stewart and D.~Wands,
  \emph{{False vacuum inflation with Einstein gravity}},
  \href{http://dx.doi.org/10.1103/PhysRevD.49.6410}{\emph{Phys. Rev.} {\bf D49}
  (1994) 6410--6433}, [\href{https://arxiv.org/abs/astro-ph/9401011}{{\tt
  astro-ph/9401011}}].

\bibitem{Dine:1983ys}
M.~Dine, W.~Fischler and D.~Nemeschansky, \emph{{Solution of the Entropy Crisis
  of Supersymmetric Theories}},
  \href{http://dx.doi.org/10.1016/0370-2693(84)91174-2}{\emph{Phys. Lett.} {\bf
  B136} (1984) 169--174}.

\bibitem{Bertolami:1987xb}
O.~Bertolami and G.~G. Ross, \emph{{Inflation as a Cure for the Cosmological
  Problems of Superstring Models With Intermediate Scale Breaking}},
  \href{http://dx.doi.org/10.1016/0370-2693(87)90431-X}{\emph{Phys. Lett.} {\bf
  B183} (1987) 163--168}.

\bibitem{Dine:1995uk}
M.~Dine, L.~Randall and S.~D. Thomas, \emph{{Supersymmetry breaking in the
  early universe}},
  \href{http://dx.doi.org/10.1103/PhysRevLett.75.398}{\emph{Phys.Rev.Lett.}
  {\bf 75} (1995) 398--401}, [\href{https://arxiv.org/abs/hep-ph/9503303}{{\tt
  hep-ph/9503303}}].

\bibitem{Linde:1996gt}
A.~D. Linde and V.~F. Mukhanov, \emph{{Nongaussian isocurvature perturbations
  from inflation}},
  \href{http://dx.doi.org/10.1103/PhysRevD.56.R535}{\emph{Phys. Rev.} {\bf D56}
  (1997) 535--539}, [\href{https://arxiv.org/abs/astro-ph/9610219}{{\tt
  astro-ph/9610219}}].

\bibitem{Weinberg:2004kf}
S.~Weinberg, \emph{{Must cosmological perturbations remain non-adiabatic after
  multi-field inflation?}},
  \href{http://dx.doi.org/10.1103/PhysRevD.70.083522}{\emph{Phys. Rev.} {\bf
  D70} (2004) 083522}, [\href{https://arxiv.org/abs/astro-ph/0405397}{{\tt
  astro-ph/0405397}}].

\bibitem{Takeuchi:2013hza}
Y.~Takeuchi and S.~Chongchitnan, \emph{{Constraining isocurvature perturbations
  with the 21cm emission from minihaloes}},
  \href{https://arxiv.org/abs/1311.2585}{{\tt 1311.2585}}.

\bibitem{Dent:2012ne}
J.~B. Dent, D.~A. Easson and H.~Tashiro, \emph{{Cosmological constraints from
  CMB distortion}},
  \href{http://dx.doi.org/10.1103/PhysRevD.86.023514}{\emph{Phys.Rev.} {\bf
  D86} (2012) 023514}, [\href{https://arxiv.org/abs/1202.6066}{{\tt
  1202.6066}}].

\bibitem{Chluba:2013dna}
J.~Chluba and D.~Grin, \emph{{CMB spectral distortions from small-scale
  isocurvature fluctuations}},
  \href{http://dx.doi.org/10.1093/mnras/stt1129}{\emph{Mon.Not.Roy.Astron.Soc.}
  {\bf 434} (2013) 1619--1635}, [\href{https://arxiv.org/abs/1304.4596}{{\tt
  1304.4596}}].

\bibitem{Sekiguchi:2013lma}
T.~Sekiguchi, H.~Tashiro, J.~Silk and N.~Sugiyama, \emph{{Cosmological
  signatures of tilted isocurvature perturbations: reionization and 21cm
  fluctuations}},
  \href{http://dx.doi.org/10.1088/1475-7516/2014/03/001}{\emph{JCAP} {\bf 1403}
  (2014) 001}, [\href{https://arxiv.org/abs/1311.3294}{{\tt 1311.3294}}].

\bibitem{Chen:2009zp}
X.~Chen and Y.~Wang, \emph{{Quasi-Single Field Inflation and
  Non-Gaussianities}},
  \href{http://dx.doi.org/10.1088/1475-7516/2010/04/027}{\emph{JCAP} {\bf 1004}
  (2010) 027}, [\href{https://arxiv.org/abs/0911.3380}{{\tt 0911.3380}}].

\bibitem{Craig:2014rta}
N.~Craig and D.~Green, \emph{{Testing Split Supersymmetry with Inflation}},
  \href{http://dx.doi.org/10.1007/JHEP07(2014)102}{\emph{JHEP} {\bf 07} (2014)
  102}, [\href{https://arxiv.org/abs/1403.7193}{{\tt 1403.7193}}].

\bibitem{Arkani-Hamed:2015bza}
N.~Arkani-Hamed and J.~Maldacena, \emph{{Cosmological Collider Physics}},
  \href{https://arxiv.org/abs/1503.08043}{{\tt 1503.08043}}.

\bibitem{Dimastrogiovanni:2015pla}
E.~Dimastrogiovanni, M.~Fasiello and M.~Kamionkowski, \emph{{Imprints of
  Massive Primordial Fields on Large-Scale Structure}},
  \href{http://dx.doi.org/10.1088/1475-7516/2016/02/017}{\emph{JCAP} {\bf 1602}
  (2016) 017}, [\href{https://arxiv.org/abs/1504.05993}{{\tt 1504.05993}}].

\bibitem{Chung:2015tha}
D.~J.~H. Chung, \emph{{Large blue isocurvature spectral index signals
  time-dependent mass}},
  \href{http://dx.doi.org/10.1103/PhysRevD.94.043524}{\emph{Phys. Rev.} {\bf
  D94} (2016) 043524}, [\href{https://arxiv.org/abs/1509.05850}{{\tt
  1509.05850}}].

\bibitem{Peccei:1977hh}
R.~Peccei and H.~R. Quinn, \emph{{CP Conservation in the Presence of
  Instantons}},
  \href{http://dx.doi.org/10.1103/PhysRevLett.38.1440}{\emph{Phys.Rev.Lett.}
  {\bf 38} (1977) 1440--1443}.

\bibitem{Weinberg:1977ma}
S.~Weinberg, \emph{{A New Light Boson?}},
  \href{http://dx.doi.org/10.1103/PhysRevLett.40.223}{\emph{Phys.Rev.Lett.}
  {\bf 40} (1978) 223--226}.

\bibitem{Wilczek:1977pj}
F.~Wilczek, \emph{{Problem of Strong p and t Invariance in the Presence of
  Instantons}},
  \href{http://dx.doi.org/10.1103/PhysRevLett.40.279}{\emph{Phys.Rev.Lett.}
  {\bf 40} (1978) 279--282}.

\bibitem{Kim:1979if}
J.~E. Kim, \emph{{Weak Interaction Singlet and Strong CP Invariance}},
  \href{http://dx.doi.org/10.1103/PhysRevLett.43.103}{\emph{Phys. Rev. Lett.}
  {\bf 43} (1979) 103}.

\bibitem{Shifman:1979if}
M.~A. Shifman, A.~I. Vainshtein and V.~I. Zakharov, \emph{{Can Confinement
  Ensure Natural CP Invariance of Strong Interactions?}},
  \href{http://dx.doi.org/10.1016/0550-3213(80)90209-6}{\emph{Nucl. Phys.} {\bf
  B166} (1980) 493--506}.

\bibitem{Zhitnitsky:1980tq}
A.~R. Zhitnitsky, \emph{{On Possible Suppression of the Axion Hadron
  Interactions. (In Russian)}}, {\emph{Sov. J. Nucl. Phys.} {\bf 31} (1980)
  260}.

\bibitem{Dine:1981rt}
M.~Dine, W.~Fischler and M.~Srednicki, \emph{{A Simple Solution to the Strong
  CP Problem with a Harmless Axion}},
  \href{http://dx.doi.org/10.1016/0370-2693(81)90590-6}{\emph{Phys. Lett.} {\bf
  B104} (1981) 199--202}.

\bibitem{Kim:2008hd}
J.~E. Kim and G.~Carosi, \emph{{Axions and the Strong CP Problem}},
  \href{http://dx.doi.org/10.1103/RevModPhys.82.557}{\emph{Rev. Mod. Phys.}
  {\bf 82} (2010) 557--602}, [\href{https://arxiv.org/abs/0807.3125}{{\tt
  0807.3125}}].

\bibitem{Kasuya:2009up}
S.~Kasuya and M.~Kawasaki, \emph{{Axion isocurvature fluctuations with
  extremely blue spectrum}},
  \href{http://dx.doi.org/10.1103/PhysRevD.80.023516}{\emph{Phys.Rev.} {\bf
  D80} (2009) 023516}, [\href{https://arxiv.org/abs/0904.3800}{{\tt
  0904.3800}}].

\bibitem{Raffelt:2006cw}
G.~G. Raffelt, \emph{{Astrophysical axion bounds}},
  \href{http://dx.doi.org/10.1007/978-3-540-73518-2_3}{\emph{Lect. Notes Phys.}
  {\bf 741} (2008) 51--71}, [\href{https://arxiv.org/abs/hep-ph/0611350}{{\tt
  hep-ph/0611350}}].

\bibitem{Sikivie:2006ni}
P.~Sikivie, \emph{{Axion Cosmology}},
  \href{http://dx.doi.org/10.1007/978-3-540-73518-2_2}{\emph{Lect.Notes Phys.}
  {\bf 741} (2008) 19--50}, [\href{https://arxiv.org/abs/astro-ph/0610440}{{\tt
  astro-ph/0610440}}].

\bibitem{Graham:2015ouw}
P.~W. Graham, I.~G. Irastorza, S.~K. Lamoreaux, A.~Lindner and K.~A. van
  Bibber, \emph{{Experimental Searches for the Axion and Axion-Like
  Particles}},
  \href{http://dx.doi.org/10.1146/annurev-nucl-102014-022120}{\emph{Ann. Rev.
  Nucl. Part. Sci.} {\bf 65} (2015) 485--514},
  [\href{https://arxiv.org/abs/1602.00039}{{\tt 1602.00039}}].

\bibitem{Chung:2015pga}
D.~J.~H. Chung and H.~Yoo, \emph{{Elementary Theorems Regarding Blue
  Isocurvature Perturbations}},
  \href{http://dx.doi.org/10.1103/PhysRevD.91.083530}{\emph{Phys. Rev.} {\bf
  D91} (2015) 083530}, [\href{https://arxiv.org/abs/1501.05618}{{\tt
  1501.05618}}].

\bibitem{Kawasaki:2013ae}
M.~Kawasaki and K.~Nakayama, \emph{{Axions: Theory and Cosmological Role}},
  \href{http://dx.doi.org/10.1146/annurev-nucl-102212-170536}{\emph{Ann.Rev.Nucl.Part.Sci.}
  {\bf 63} (2013) 69--95}, [\href{https://arxiv.org/abs/1301.1123}{{\tt
  1301.1123}}].

\bibitem{Chen:2015dga}
X.~Chen, M.~H. Namjoo and Y.~Wang, \emph{{On the equation-of-motion versus
  in-in approach in cosmological perturbation theory}},
  \href{http://dx.doi.org/10.1088/1475-7516/2016/01/022}{\emph{JCAP} {\bf 1601}
  (2016) 022}, [\href{https://arxiv.org/abs/1505.03955}{{\tt 1505.03955}}].

\bibitem{CLN}
``Class library for numbers.'' \url{http://www.ginac.de/CLN/}.

\end{thebibliography}\endgroup

\end{document}